\PassOptionsToPackage{prologue,dvipsnames}{xcolor}
\documentclass{article}
\usepackage{fullpage}
\usepackage{microtype}
\usepackage{graphicx}
\usepackage{subcaption}
\usepackage{multirow}
\usepackage{placeins}
\usepackage{booktabs, colortbl} 
\usepackage[square]{natbib}
\usepackage[colorlinks=true, linkcolor=BrickRed, citecolor=darkgreen, urlcolor=blue]{hyperref}

\usepackage{comment}

\let\oldemph\emph
\renewcommand{\emph}[1]{{\color{BrickRed}\oldemph{#1}}}

\usepackage{wrapfig}

\usepackage{amsmath}
\usepackage{amssymb}
\usepackage{mathtools}
\usepackage{amsthm}

\usepackage[capitalize,noabbrev,nameinlink]{cleveref}

\theoremstyle{definition}

\theoremstyle{remark}

\usepackage[textsize=tiny]{todonotes}

\usepackage{booktabs}       
\usepackage{nicefrac}       
\usepackage{microtype}      
\usepackage{xcolor}  
\definecolor{winered}{rgb}{0.5,0.1,0.1}
\definecolor{darkgreen}{rgb}{0,0.35,0}
       
\usepackage{xspace}

\usepackage[T1]{fontenc}

\usepackage{tikz}
\usetikzlibrary{patterns}
\usetikzlibrary{calc}
\usetikzlibrary{arrows.meta}
\usetikzlibrary{positioning}
\usetikzlibrary{shapes.geometric}
\usetikzlibrary{decorations.pathreplacing}
\usepackage{pgfplots}

\usepackage{amsmath}

\DeclareMathOperator{\prm}{\mathsf{PR}}
\DeclareMathOperator{\katz}{\mathsf{K}}

\usepackage{mathtools}

\newcommand{\absorbkatz}{{\textsc{AbsorbKatz}}\xspace}

\newcommand{\mesrank}{{\textsc{MesRank}}\xspace}
\newcommand{\meskatz}{{\textsc{MesKatz}}\xspace}

\newcommand{\bosrank}{{\textsc{BosRank}}\xspace}
\newcommand{\boskatz}{{\textsc{BosKatz}}\xspace}

\newcommand{\toprank}{{\textsc{TopRank}}\xspace}
\newcommand{\topkatz}{{\textsc{TopKatz}}\xspace}

\renewcommand{\top}{{\textsc{Top}}\xspace}
\newcommand{\bos}{{\textsc{Bos}}\xspace}
\newcommand{\mes}{{\textsc{Mes}}\xspace}
\newcommand{\random}{{\textsc{Random}}\xspace}
\newcommand{\tim}{{\textsc{Tim}+}\xspace}

\newcommand{\Succ}{{\mathrm{Succ}}}
\newcommand{\Pred}{\mathrm{Pred}}

\usepackage{pifont}
\definecolor{crimson}{RGB}{220, 20, 60}
\definecolor{darkgreen}{rgb}{0,0.5,0}

\definecolor{mossgreen}{rgb}{0.68, 0.87, 0.68}
\definecolor{lightpink}{rgb}{1.0, 0.71, 0.76}
\definecolor{lightgray}{rgb}{0.83, 0.83, 0.83}

\usepackage{txfonts}

\usepackage{amsthm}

\usepackage{cleveref}

\newcommand{\Was}{W\k{a}s}

  \tikzset{     
    e4c node/.style={circle,draw,minimum size=0.3cm,inner sep=0,font=\tiny},
    e4b node/.style={circle,draw,minimum size=0.8cm,inner sep=0,font=\normalsize},
    selected/.style={draw=BrickRed,double}, 
    selected2/.style={},
    selected3/.style={pattern=north east lines, pattern color=NavyBlue}, 
    e4c edge/.style={sloped,above,font=\footnotesize,-{Classical TikZ Rightarrow[length=0.75mm]},draw,thin},
    e4c path/.style={-{Classical TikZ Rightarrow[length=0.75mm]},draw,thin}
  }
\newcommand{\naturals}{{{\mathbb{N}}}}
\newcommand{\reals}{{{\mathbb{R}}}}

\makeatletter
               {\list{}{%
                   \setlength{\labelwidth}{\z@}%
                   \setlength{\labelsep}{1pt}
                   \setlength{\leftmargin}{0pt}
                   \setlength{\itemsep}{2pt}
                   \setlength{\parsep}{2pt}
                   \setlength{\topsep}{2pt}
                   \setlength{\parskip}{2pt}
                   \setlength{\itemindent}{\z@}%
                   }}%
               {\endlist}
\makeatother

\title{Proportional Selection in Networks}

\usepackage{authblk}

\date{}

\author[1]{Georgios Papasotiropoulos}
\author[1]{Oskar Skibski}
\author[1]{Piotr Skowron}
\author[2]{Tomasz \Was}

\affil[1]{{\small{University of Warsaw}}}
\affil[2]{{\small{University of Oxford}}}

\begin{document}

\maketitle
	\begin{abstract}
	We address the problem of selecting $k$ representative nodes from a network, aiming to simultaneously achieve two objectives: identifying the most influential nodes and ensuring that the selection proportionally reflects the diversity within the network.
	We propose a general approach to accomplish this by combining ideas from network science and computational social choice. Notably, our algorithms depend only on the connections between nodes and do not utilize any additional information that would explicitly identify groups of nodes.
	We analyze them theoretically, and demonstrate their effectiveness through a series of experiments.
\end{abstract}

	\section{Introduction}\label{sec:intro}

Consider the \textit{peer-selection problem} of selecting $k$ nodes from a network.
Our goal is twofold: to identify the most influential nodes,
and to ensure that the selection proportionally represents the structure of the network. 
For instance, consider a network composed of three densely connected groups,
containing 50\%, 30\%, and 20\% of all nodes, respectively,
with relatively sparse connections between the groups. 
If the objective is to select $k = 10$ nodes,
a proportional approach would involve
selecting five most influential nodes from the first group,
three from the second, and two from the third group. 
In this paper we design selection methods that capture this intuition,
yet apply to networks with arbitrary structures.

In recent years,
a number of papers proposed methods to
select influential nodes in a network
in a way that guarantees fair treatment
of different communities~\citep{anwar2021balanced,stoica2020seeding,tsang2019groupfairness,tsioutsiouliklis2021fairness}.
All of these focus on the so-called \emph{feature-aware} fairness,
which means that the guarantees
are given based on specific attributes shared by a group.
This may be problematic in some applications,
as the characteristics of the participants are not always available, e.g.,
due to anonymity~\citep{kilbertus2018blind}.
At other times, such attributes cannot be used
due to ethical, legal, or privacy reasons~\citep{van2023using},
or can be incomplete, unreliable, or noisy~\citep{ashurst2023fairness}.
In social-network settings, groups may naturally emerge dynamically 
or implicitly from structural relations rather than from explicit features.
Hence, in this work we focus on \emph{feature-blind} fairness,
which gives proportional guarantees to the groups of nodes
only based on the topology of the graph.
A recent survey on fairness in social network analysis
underscores the need for developing methods with such guarantees~\citep{saxena2024fairsna}.\looseness-1

In order to achieve this goal,
we bridge together two distinct areas of research:
\emph{centrality measures} that aim at finding the most influential
elements of the network~\citep{koschutzki2005centrality};
and \emph{social choice theory} which focuses, among others,
on the selection of proportional outcomes~\citep{lac-sko:b:approval-survey}.
We propose a framework that transforms
a centrality measure into an election
to which we apply state-of-the-art methods from social choice
that exhibit strong theoretical proportionality guarantees~\citep{PetPieSko-2021-EqualSharesPB,papasotiropoulos2024method}.
While we focus on certain prominent centrality measures, this is for illustration purposes only; 
our approach is designed to be broad and general enough
to be compatible with most centralities,
as well as with custom machine learning models for assessing node importance.
As a result, our work enables the extension of
existing, well-established metrics of importance,
to the task of selecting groups of nodes,
in a way that maintains the underlying logic of each metric
while also ensuring proportional representation.

We analyze our methods axiomatically and characterize their behavior on graph classes, where proportional solutions can be intuitively captured.
Our theoretical results suggest that the proposed methods can identify and fairly represent large cohesive groups in any input, without requiring node labels or agent attributes.
Our experiments highlight the effectiveness of the proposed methods in promoting proportional outcomes,
while still finding highly influential nodes.
In the experiments,
we rely on structured graphs where proportionality is both interpretable and measurable,
and show that our methods achieve fair representation across different communities---even when membership information is not explicitly provided and can only be inferred from the network’s structure.

Our framework and approach have broad applicability across various real-world scenarios.
Imagine a network where nodes represent political blogs or news websites,
and links indicate references between them.
In this context, proportional selection ensures a balanced representation of major viewpoints---such as left-wing and right-wing political perspectives,
or other widely held opinions that we may not even be aware of---within the selected set of sites.
Similarly, one may use our methods to find influential users
in social media discussions,
while giving a fair treatment to all discussion circles;
or select prominent researchers in an area
ensuring that all subareas are fairly recognized. 
In online decision-making platforms where participants may delegate their voting rights to others, such as LiquidFeedback~\citep{behrens2014principles} or Google Votes~\citep{hardt2015google}, our methods can identify the individuals that proportionally represent the electorate.

\smallskip
Proofs and additional experimental results
can be found in the Appendix.

\subsection*{Further Related Work}
Computational social choice distinguishes between three major objectives in the context of selecting subsets of candidates: individual excellence, diversity, and proportionality~\citep{FSST-trends,lac-sko:b:approval-survey}.
Individual excellence aims to select the best candidates overall.
This corresponds to choosing the $k$ nodes with the highest centrality,
which can severely neglect minority groups. 
At the other extreme, diversity ensures that as many individuals as possible enjoy some representation.
Group centrality measures that promote diversity
have been already explored \citep{lyu2021centrality,mavroforakis2015absorbing,kempe2003maximizing,lin2010improving}.
However, under this objective,
large groups may be significantly underrepresented
in favor of extremely small groups of disconnected nodes.
We take an intermediate approach,
aiming to represent different components of a network
at a rate proportional to their size.\looseness-1

An active line of research applies the idea of proportional representation
beyond traditional voting setting.
It was considered in 
the contexts of online deliberative discourse~\citep{revel2025representative},
AI-augmented civic participation platforms~\citep{fish2023generative},
fairness-aware versions of the secretary problem~\citep{papasotiropoulosfairness},
and ranking aggregation settings relevant to online e-commerce platforms~\citep{lederer2024squared}.
Our work relates to the research on group centrality measures~\citep{everett1999centrality,angriman2020group},
but proportionality has not been yet considered in this context.
Strategic aspects of node selection in networks have also been explored~\citep{alon2011sum,holzman2013impartial,aziz2016strategyproof}.
Finally, there is a long line of work on community detection in networks,
node representation learning, and facility location---indicatively see,
respectively,~\citep{chunaev2020community,bothorel2015clustering},
\citep{shi2024label,kose2021fairness}, and~\citep{zhou2022strategyproof,blanco2023fairness,lam2024proportional}.
However, in contrast to what is typically done in that research,
our focus is on ensuring that each group of nodes is proportionally represented
without relying on additional attributes of the nodes.

\section{Preliminaries}
\label{sec:prelims}
We consider unweighted simple directed graphs.
Yet, our approach can be adapted to take into account weighted edges.
A \emph{graph} (or a \emph{network}) is a pair, $G = (V,E)$, where~$V$ is a set of $n$ nodes and $E \subseteq V \times V$ is a set of $m$ edges; the edges are ordered pairs of nodes.
An edge $(u,v)$ is an \emph{outgoing edge} for node $u$ and an \emph{incoming edge} for $v$.
The number of outgoing (resp., incoming) edges of $u$ is called \emph{out-degree} (resp., \emph{in-degree}) of $u$ and denoted by $\deg^+(u)$ (resp., $\deg^-(u)$).\looseness-1

A \emph{walk} is a sequence of nodes $(v_1,\dots,v_k)$ such that every two consecutive nodes are connected by an edge: $(v_i,v_{i+1}) \in E$ for every $i \in \{1,\dots,k-1\}$.
The length of such a walk is equal to $k-1$, i.e., the number of edges in the walk.
Note that a single node is a walk of length 0.
A node $u$ is called a \emph{predecessor} of $w$ if there exists a walk $(v_1,\dots,v_k)$ such that $u = v_1$ and $w = v_k$.
A node $u$ is a \emph{successor} of $w$ if $w$ is a predecessor of $u$.
The set of predecessors of node $u$ is denoted by $\Pred(u)$ and its set of successors is denoted by $\Succ(u)$.
For a set of nodes $S \subseteq V$ we define 
$\Succ(S) = \bigcup_{u \in S} \Succ(u)$.
The set of all walks in $G$ is denoted by $\Omega(G)$.

For a set of nodes $S \subseteq V$, the subgraph induced by $S$, denoted by $G[S]$, is the graph $(S,\{(u,v) \in E: u,v \in S\})$, i.e., the graph containing $S$ and the edges between nodes from $S$.
A graph $G$ is \emph{strongly connected} if for every two nodes $u,v$ of $G$ there exists a walk from $u$ to $v$; $G$ is \emph{weakly connected} if every two nodes are connected by a walk in the underlying undirected graph inferred by $G$. 
A \emph{component} of $G$ is a maximal weakly connected subgraph of $G$.
A \emph{clique} is a graph in which, for every two nodes $u,v,$ there exist edges $(u, v)$ and $(v, u)$.

A graph $G=(V,E)$ is called (directed) \emph{bipartite} if all walks in $\Omega(G)$ are of length at most $1$, in other words, if its set of nodes $V$ can be divided into two disjoint sets $V = V_1 \cup V_2$ such that every edge in $E$ is an outgoing for a node in $V_1$ and an incoming for a node in $V_2$.
A graph $G=(V,E)$ is called \emph{functional} if $\deg^+(u)\leq 1,$ for all nodes $u \in V$.

Given a network $G(V, E)$ and an integer $k<n$ our goal is to select $k$ nodes from $V$\!. A method that performs this selection is referred to as a \emph{group selection rule for networks}, in short, a \emph{rule}, and denoted by $\mathcal{R}$. We will also refer to the outcome of such a rule simply as $\mathcal{R}(G, k)$.

\subsection{Centrality Measures}
\label{sec:centralities}
A \emph{centrality measure} $F$ is a function that for each graph $G=(V,E)$  and node $v \in V$ assigns a real value, denoted by $F_G(v)$. The higher the value, the more central the node is considered.
Our methods can be combined with any centrality measure; for illustration, we focus on the most popular centrality measures based on walks; PageRank and Katz centrality.

\smallskip
\textbf{PageRank~\citep{PagBriMotWin-1999-PageRank}}: For a given decay factor $\alpha \in (0,1)$, PageRank of a node $v$ in graph $G$ is defined as:
\begin{equation}\label{eq:pagerank}
	\prm^{\alpha}_G(v) = \sum_{(u_1,\dots,u_t,v) \in \Omega(G)} \frac{\alpha^t}{\prod_{i=1}^{t} \deg^+(u_i)}.
\end{equation}

\noindent At a high level, PageRank of $v$ is proportional to the expected number of times that $v$ is being visited by a random walk that starts from a random node and in each step follows an outgoing edge chosen uniformly at random or ends the walk with a probability of $1-\alpha$.
Hence, it mostly depends on the number and the importance of its predecessors.
We note that many variants of it appear in the literature~\citep{WasSki-2023-PageRank}.\looseness-1

\smallskip
\textbf{Katz centrality~\citep{Kat-1953-KatzCentrality}}:
For a given decay factor $\alpha\in (0,\nicefrac{1}{\lambda})$, where $\lambda$ is the largest eigenvalue of the adjacency matrix of $G$, Katz centrality of a node $v$ in $G$ is defined as:%
\begin{equation}\label{eq:katz}
	\katz^{\alpha}_G(v) = \sum_{(u_1,\dots,u_t,v) \in \Omega(G)} \alpha^t.
\end{equation}

\noindent It can be considered a parallel version of PageRank, where the importance of a node is not split between its outgoing edges, but transferred through all of these simultaneously.

For notational simplicity, when it is clear from the context, we omit $G$ and $\alpha$ from the scripts. 

\subsection{Election Rules}
\label{sec:comsoc_rules}
A committee election is a triple $\left(V, C, \mu = \{\mu_i\}_{i \in V}\right)$, where $\mu_i\colon C \to \reals_{+}$ is a function representing the preferences of a voter $i \in V$ over the candidates in $C$. A higher value of $\mu_i(c)$ indicates a stronger level of support from $i$ towards $c$.
A voting rule is a function that for each election $(V,C,\mu)$ and natural number $k \in \naturals$ returns a subset of $k$ candidates from $C$. Voters' preferences can be expressed in various ways, e.g., using approval ballots where $\mu_i(c) = 1$ if voter~$i$ approves candidate $c$ and $\mu_i(c) = 0$, otherwise, or using general utility functions in which $\mu_i(c)$ is an arbitrary non-negative value indicating the satisfaction of $i$ from electing $c$. We particularly focus on scenarios where $V \equiv C$, i.e., when the voters aim to select a committee from among themselves.

Two of the simplest election rules are \emph{Approval Voting} (AV) and \emph{Satisfaction Approval Voting} (SAV). 
Adapting their principles to our setting, we will say that each candidate $c$ receives a score from each voter $i \in V$, which equals $\mu_i(c)$ for AV and $\mu_i(c)/\sum_{c'\in C}{\mu_i(c')}$ for SAV. In both, the $k$ candidates having the maximum total score form the winning committee.
A more involved voting rule that aims to select sets of candidates in a proportional way has been proposed under the name Method of Equal Shares (MES)~\citep{Peters:Skowron:2020,PetPieSko-2021-EqualSharesPB}:
Let $b_i$ be a virtual budget of a voter $i$, initially set to $\nicefrac{k}{n}$. 
The rule operates in rounds. We say that a not yet elected candidate $c$ is \emph{$\rho$-affordable} for $\rho \in \mathbb{R}_{+}$, if its supporters can cover its (unit) cost in such a way that each of them pays $\rho$ per unit of utility, or all of their remaining funds, i.e., if $\sum_{i \in V} \min\left(b_i, \mu_i(c) \cdot\rho\right)\geq 1$. We calculate the minimum value of $\rho$ satisfying $\sum_{i \in V} \min\left(b_i, \mu_i(c) \cdot\rho\right)\geq 1$.
In a given round the method selects the candidate that is $\rho$-affordable for the lowest possible value of $\rho$ and updates the budgets accordingly: $b_i := b_i - \min\left(b_i, \mu_i(c) \cdot\rho\right)$.
The procedure stops if there is no $\rho$-affordable candidate for any finite $\rho$.
Note that the method may terminate with less than $k$ candidates selected. In this case, in our experiments, we will use the method with the Add1U completion~\citep{fal-fli-pet-pie-sko-sto-szu-tal:pb-experiments}. Additionally, the very recently proposed \emph{Method of Equal Shares with Bounded Overspending} (BOS)~\citep{papasotiropoulos2024method}---a robust variant of MES that balances proportionality and efficiency---will also be explored in our simulations.\looseness-1

\section{Framework for Election-Based Selection}
\label{sec:rules}

A natural approach to selecting influential nodes in a network is to choose those with the highest centrality. In the context of liquid democracy, \citet{boldi2011viscous} proposed selecting the $k$ nodes with the highest PageRank---a method we refer to as \toprank. This use of PageRank captures the diminishing trust along delegation paths to selected nodes.
We define \topkatz\ as the analogous rule based on Katz centrality.
As we will show, these methods can critically fail to ensure proportional representation in the network. While proportionality has been highlighted as a key open problem in liquid democracy~\citep{brill2018interactive}, no proportional selection rules have been proposed to date, to the best of our knowledge.
To address this, we introduce a general framework for proportional selection that applies not only to liquid democracy (see the recent survey by \citet{liquid_survey} for details on the topic) but also to a broader range of network settings. Our methods take only graphs as input, without any additional information that would explicitly identify groups of nodes within the network.
This approach constitutes our main conceptual contribution.

Our proposal leverages proportional committee election rules. At a high level, it transforms the input graph into an election based on nodes' importance.
By interpreting
nodes as participants, it
applies a suitable rule to select representatives. In particular, for any two distinct nodes $u$ and $v$, we define a utility for node $u$ derived from including node $v$ in the selected set.
Such an assessment can be derived from most centrality measures in a natural way, but can also be the result of a link prediction, similarity measures, or other custom machine learning models.
For PageRank, with a decay factor $\alpha$, we define $\mu^{\alpha}_G(u,v)$ as the expected number of visits at $v$ of the random walk that starts at node $u$, based on \cref{eq:pagerank}:
\begin{equation}\label{eq:pagerank_evaluations}
	\mu^{\alpha}_G(u,v):= \sum_{(u_1,u_2,\dots,u_k,v) \in \Omega(G) : u = u_1} \frac{\alpha^k}{\prod_{i=1}^{k} \deg^+(u_i)}.
\end{equation}
According to \Cref{eq:pagerank_evaluations}, the utility of $u$ from selecting $v$ is high if node $v$ can be reached with high probability from $u$.
If we interpret edges as votes' delegation, the node to which a vote can be delegated more directly
is preferred. 
Note that $\prm^\alpha_G(v) = \sum_{u \in V} \mu^{\alpha}_G(u,v)$,
so \toprank is equivalent to AV rule applied to the election $(V,V,\mu^\alpha_G)$.
Instead, we mainly use the Method of Equal Shares due to its strong proportionality guarantees,
resulting in a rule we refer to as \mesrank. The proposed selection rule is parameterized by a value $\alpha \in (0,1)$, as it relies on \cref{eq:pagerank}, and essentially applies MES to the election where the set of candidates coincides with the set of voters $V$, and preferences are determined by \Cref{eq:pagerank_evaluations}. In short,
$\mesrank^{\alpha}(G,k) := \textsc{MES}((V,V, \mu^{\alpha}_G), k).$
For Katz centrality the definition of a utility function is analogous---but based on \cref{eq:katz}---and this gives rise to a rule that we will call \meskatz.

\section{Theoretical Analysis}\label{sec:case-studies}
In this section, we provide a theoretical analysis of \mesrank and \meskatz, beginning with the structural analysis of their behaviour based on specific graph classes and then turning to establishing axiomatic properties satisfied by the methods.\looseness-1

In particular, we first illustrate our methods on
bipartite and functional graphs, as defined in \Cref{sec:prelims}.
In these graph families, proportionality can be intuitively captured, and thus they provide a first way of examining how the proposed methods achieve proportionality and how they differ from one another.
Bipartite graphs mimic the scenario of representative democracy,
where the set of candidates is separate from the set of voters.
Functional graphs can be viewed as elections with 1-approval ballots 
(where each voter supports at most one candidate). They also reflect the delegation structure used in LiquidFeedback~\citep{behrens2014principles}---%
the most prominent implementation of liquid democracy.
Functional graphs were also the exclusive focus of \citet{boldi2011viscous} where \toprank was proposed.

\begin{figure}[t!]
	\centering
	\begin{minipage}[t]{0.48\textwidth}
		\centering
		\begin{tikzpicture}[x=0.55cm,y=1cm] 
			\tikzset{     
				e4c node/.style={circle,draw,minimum size=0.3cm,inner sep=0,font=\footnotesize}, 
				e4c edge/.style={sloped,above,font=\footnotesize}
			}
			\node[e4c node,draw=none] (14) at (-0.5, 0.00) {$V_1\!:$};
			\node[e4c node,draw=none] (15) at (-0.5, 1.00) {$V_2\!:$};
			\node[e4c node] (1) at (1.00, 0.00) {}; 
			\node[e4c node] (2) at (2.00, 0.00) {}; 
			\node[e4c node] (3) at (3.00, 0.00) {}; 
			\node[e4c node,draw=none] (4) at (4.00, 0.00) {\ $\dots$}; 
			\node[e4c node] (5) at (5.00, 0.00) {}; 
			\node[e4c node] (6) at (6.00, 0.00) {}; 
			\node[e4c node] (7) at (7.00, 0.00) {}; 
			\node[e4c node,draw=none] (8) at (8.00, 0.00) {\ $\dots$}; 
			\node[e4c node] (9) at (9.00, 0.00) {}; 
			\node[e4c node] (10) at (10.00, 0.00) {}; 
			\node[e4c node] (11) at (11.00, 0.00) {}; 
			\node[e4c node,draw=none] (12) at (12.00, 0.00) {\ $\dots$}; 
			\node[e4c node] (13) at (13.00, 0.00) {}; 
			
			\node[e4c node,selected,selected2,selected3] (c1) at (1.50, 1.00) {}; 
			\node[e4c node,selected,selected2,selected3] (c2) at (2.50, 1.00) {}; 
			\node[e4c node,draw=none] (c123) at (3.50, 1.00) {\ $\dots$}; 
			\node[e4c node,selected,selected2] (c3) at (4.50, 1.00) {};
			
			\node[e4c node,selected3] (c4) at (6.00, 1.00) {}; 
			\node[e4c node] (c5) at (7.00, 1.00) {}; 
			\node[e4c node,draw=none] (c456) at (8.00, 1.00) {\ $\dots$}; 
			\node[e4c node] (c6) at (9.00, 1.00) {}; 
			
			\node[e4c node,selected3] (c7) at (10.00, 1.00) {}; 
			\node[e4c node] (c8) at (11.00, 1.00) {}; 
			\node[e4c node,draw=none] (c789) at (12.00, 1.00) {\ $\dots$}; 
			\node[e4c node] (c9) at (13.00, 1.00) {}; 
			
			\draw [decorate,decoration={brace,amplitude=5pt,mirror,raise=1.5ex}]
			(0.6,0) -- (5.4,0) node[midway,yshift=-20]{\small $40\%$};
			\draw [decorate,decoration={brace,amplitude=5pt,mirror,raise=1.5ex}]
			(5.6,0) -- (9.4,0) node[midway,yshift=-20]{\small $30\%$};
			\draw [decorate,decoration={brace,amplitude=5pt,mirror,raise=1.5ex}]
			(9.6,0) -- (13.4,0) node[midway,yshift=-20]{\small $30\%$};
			
			\draw [decorate,decoration={brace,amplitude=5pt,raise=1.5ex}]
			(1.1,1) -- (4.9,1) node[midway,yshift=20]{\small $k$};
			\draw [decorate,decoration={brace,amplitude=5pt,raise=1.5ex}]
			(5.6,1) -- (9.4,1) node[midway,yshift=20]{\small $k$};
			\draw [decorate,decoration={brace,amplitude=5pt,raise=1.5ex}]
			(9.6,1) -- (13.4,1) node[midway,yshift=20]{\small $k$};
			
			\path[e4c path]
			(1) edge[e4c edge]  (c1)
			(2) edge[e4c edge]  (c1)
			(3) edge[e4c edge]  (c1)
			(5) edge[e4c edge]  (c1)
			(1) edge[e4c edge]  (c2)
			(2) edge[e4c edge]  (c2)
			(3) edge[e4c edge]  (c2)
			(5) edge[e4c edge]  (c2)
			(1) edge[e4c edge]  (c3)
			(2) edge[e4c edge]  (c3)
			(3) edge[e4c edge]  (c3)
			(5) edge[e4c edge]  (c3)
			
			(6) edge[e4c edge]  (c4)
			(7) edge[e4c edge]  (c4)
			(9) edge[e4c edge]  (c4)
			(6) edge[e4c edge]  (c5)
			(7) edge[e4c edge]  (c5)
			(9) edge[e4c edge]  (c5)
			(6) edge[e4c edge]  (c6)
			(7) edge[e4c edge]  (c6)
			(9) edge[e4c edge]  (c6)
			(10) edge[e4c edge]  (c7)
			(11) edge[e4c edge]  (c7)
			(13) edge[e4c edge]  (c7)
			(10) edge[e4c edge]  (c8)
			(11) edge[e4c edge]  (c8)
			(13) edge[e4c edge]  (c8)
			(10) edge[e4c edge]  (c9)
			(11) edge[e4c edge]  (c9)
			(13) edge[e4c edge]  (c9)
			;
		\end{tikzpicture}
		\caption{%
			Selecting $k$ nodes from a bipartite graph consisting of three fully connected groups.
			\toprank (\textcolor{winered}{red} double lines) selects all $k$ nodes from the first group of $V_2$. \mesrank (\textcolor{NavyBlue}{blue} pattern) selects $0.4k$ of nodes from the first group and $0.3k$ from each other group.
		}
		\label{fig:voting}
	\end{minipage}
	\hfill
	\begin{minipage}[t]{0.48\textwidth}
		\centering
		\begin{tikzpicture}[x=6cm,y=3cm] 
			\node[e4c node,selected,selected2,selected3] (1) at (0.51, 1.50) {}; 
			\node[e4c node,selected,selected3] (2) at (0.195, 1.35) {}; 
			\node[e4c node,selected2,selected3] (3) at (0.065, 1.10) {}; 
			\node[e4c node,selected2,selected3] (4) at (0.325, 1.10) {}; 
			\node[e4c node] (5) at (0.00, 0.85) {}; 
			\node[e4c node] (6) at (0.13, 0.85) {}; 
			\node[e4c node] (7) at (0.26, 0.85) {}; 
			\node[e4c node] (8) at (0.39, 0.85) {}; 
			\node[e4c node,selected,selected3] (9) at (0.83, 1.35) {}; 
			\node[e4c node,selected,selected2,selected3] (10) at (0.99, 1.27) {}; 
			\node[e4c node,selected] (11) at (0.83, 1.19) {}; 
			\node[e4c node,selected] (12) at (0.99, 1.11) {}; 
			\node[e4c node,selected2] (13) at (0.83, 1.03) {}; 
			\node[e4c node] (14) at (0.99, 0.95) {}; 
			\node[e4c node] (15) at (0.83, 0.87) {}; 
			
			\path[e4c path]
			(15) edge[e4c edge]  (14)
			(14) edge[e4c edge]  (13)
			(2) edge[e4c edge]  (1)
			(3) edge[e4c edge]  (2)
			(4) edge[e4c edge]  (2)
			(5) edge[e4c edge]  (3)
			(6) edge[e4c edge]  (3)
			(7) edge[e4c edge]  (4)
			(8) edge[e4c edge]  (4)
			(9) edge[e4c edge]  (1)
			(10) edge[e4c edge]  (9)
			(11) edge[e4c edge]  (10)
			(12) edge[e4c edge]  (11)
			(13) edge[e4c edge]  (12)
			(14) edge[e4c edge]  (13)
			(15) edge[e4c edge]  (14)
			;
		\end{tikzpicture}
		\caption{%
			Selecting $k=6$ nodes from a directed in-tree with two equally sized branches. \toprank\ (\textcolor{winered}{red} double lines) selects $4$ nodes
			from the right branch and $1$ node from the left, even though the latter has about half the nodes of the network. \mesrank\ (\textcolor{NavyBlue}{blue} pattern) achieves a more balanced selection.
		}
		\label{fig:functional}
	\end{minipage}
\end{figure}

\subsection{Characterization for Bipartite Graphs}
\label{sec:bipartite}
We begin by looking at directed bipartite graphs that are particularly useful in highlighting the differences between methods based on PageRank and
Katz centrality.
Moreover, they are suitable for illustrating the characteristic behavior of our proportional selection rules.
Without loss of generality we assume that at least $k$ nodes have incoming edges. 
Since there are only walks of length $0$ and $1$ in the examined graphs, the Katz centrality and PageRank of each node $v$ can be easily determined. Specifically, if $v$ belongs to $V_1$, its centrality is minimal, i.e., $\prm^{\alpha}_G(v) = \katz^{\alpha}_G(v) = 1.$  
Conversely, if $v$ belongs to $V_2$, then:  
$\prm^{\alpha}_G(v) = 1 + \sum_{(u, v) \in E} \frac{\alpha}{\deg^+(u)},$ and $\katz^{\alpha}_G(v) = 1 + \alpha \cdot \deg^-_G(v).$
In words, nodes in $V_2$ get $\alpha$ per incoming edge under Katz centrality, and $\alpha$ divided by the number of outgoing edges per each supporting voter under PageRank.  
As a result,
\topkatz \textit{and} \toprank \textit{applied to a bipartite graph
	are equivalent to applying AV and SAV, respectively, 
	to the corresponding approval election instance.}
Consequently, no proportionality guarantees can be achieved by these two rules.

The approach based on the Method of Equal Shares works differently and as follows:
The utility that node $u$ assigns to node $v$ is $\mu^{\alpha}_G(u, v) = \nicefrac{\alpha}{\deg^+(u)}$ for PageRank and $\mu^{\alpha}_G(u, v) = \alpha$ for Katz centrality.
Consequently, 
\meskatz \textit{applied to a bipartite graph is equivalent to applying} MES 
\textit{to the corresponding approval election instance,
	while} \mesrank \textit{is equivalent to applying a version of} MES 
\textit{in which the utilities are divided by the number of approvals of each voter.}

For a visualization, consider the bipartite graph depicted in \Cref{fig:voting}. 
Methods that select nodes based on the highest PageRank or Katz centrality will choose all $k$ nodes from the first group of $V_2$, ignoring the votes of the remaining $60\%$ of voters. On the other hand, \mesrank and \meskatz will select $0.4k$ nodes from the first group and $0.3k$ nodes from each other group of $V_2$ (up to rounding). 
This follows from the EJR property of MES~\citep{Peters:Skowron:2020} which says that every large enough group of voters approving the same set of candidates should get a proportional representation in the selected outcome. It is straightforward to generalize this observation, based on EJR, to bipartite graphs with groups of any size.\looseness-1

\subsection{Group Selection on Functional Graphs}
We now move to functional graphs.
PageRank and Katz centralities are identical on such graphs, hence, for ease of exposition, our analysis will focus on rules based on PageRank.
For simplicity, we assume that the decay factor $\alpha$ approaches 1. Then the PageRank of nodes that do not lie on any cycle is nearly equal to its number of predecessors.

Consider an arbitrary connected functional graph.
Such a graph consists of one cycle or a root node and in-trees attached to it.
Nodes from the cycle or root clearly have the maximal PageRank and obtain non-zero (close to $1$) utility from all the nodes.
Thus, if the cycle contains at least $k$ nodes, then both \toprank and \mesrank would select only nodes from the cycle.
If not, both methods select all nodes from the cycle (or the root) and then some nodes from the attached in-trees.
This is, however, where the two methods begin to differ; we refer to \Cref{fig:functional} for a specific example.
\toprank selects nodes with the highest number of predecessors and may pick all or most nodes from the same in-tree.
In contrast, \mesrank selects nodes from each in-tree proportionally to its size, while preferring nodes supported by more predecessors.

\subsection{Group Selection on General Graphs}
\label{sec:axioms}

We now introduce formal axiomatic properties that capture the intuition that a sufficiently large and cohesive group of nodes deserves a proportional number of representatives in the selected outcome.
Our axioms are inspired by the literature on multiwinner election rules~\citep{aziz2017justified,lac-sko:b:approval-survey}.
The idea is that each node should influence the selection of roughly $\nicefrac{k}{n}$ fraction of the selected nodes.
Consequently, a cohesive group $S$ is entitled to at least $\lfloor k \cdot \nicefrac{|S|}{n} \rfloor$ representatives. 
The key question, then, is: in the context of our study, which groups of nodes can be considered cohesive, and which nodes qualify as proper representatives of these groups?
Roughly speaking, we view a group of nodes as cohesive if all nodes are connected to each other, 
either directly or indirectly.

The groups of nodes that are most cohesive
are the ones forming a component that is a clique.
Our first axiom states that each such a component
is entitled to a representation
proportional to its size.

\smallskip
\textbf{Clique-Entitlement:}  
A rule $\mathcal{R}$ satisfies Clique-Entitlement if for every graph $G = (V, E)$,  
and every component $S \subseteq V$  
that is a clique,  
it holds that $|S \cap \mathcal{R}(G, k)| \ge \lfloor k \cdot \nicefrac{|S|}{n} \rfloor$.  
\smallskip

\toprank \textit{and} \topkatz \textit{do not satisfy even this basic axiom.} Both may entirely overlook nodes from a sufficiently large clique 
when a larger component is present (regardless of whether it is densely connected or not) selecting only nodes from the latter. This again highlights that they are not suitable for proportional representation.
In contrast, \textit{this minimal guarantee is satisfied by} \mesrank \textit{and} \meskatz.

We can generalize Clique-Entitlement and require the component only being strongly connected, without necessarily being a clique.
This gives rise to the folowing axiom.

\smallskip
\textbf{Component-Entitlement:}  
A rule $\mathcal{R}$ satisfies Component-Entitlement if for every graph $G = (V, E)$,  
and every component $S \subseteq V$  
that is strongly connected,  
it holds that $|S \cap \mathcal{R}(G, k)| \ge \lfloor k \cdot \nicefrac{|S|}{n} \rfloor$.  
\smallskip

We go a step beyond and define an even stronger property.
Consider a strongly connected subgraph within a larger component of a graph.
Since this subgraph is not entirely separate from other nodes, its delegated votes may flow outside the group.
However, these votes will always flow to their successors---which are shared by all nodes in the (strongly connected) subgraph.
Under component-entitlement, we assume that a group deserving representation would be represented by its own members;
here we allow for representation through successors, as those are also supported by the members of the considered group.

\smallskip
\textbf{Subgraph-Entitlement:}
A rule $\mathcal{R}$
satisfies 
Subgraph-Entitlement if for every graph $G=(V,E)$,
and every subset of nodes $S \subseteq V$ 
such that $G[S]$ is strongly connected, it holds that
$|(S \cup \Succ(S)) \cap \mathcal{R}(G,k)| \ge \lfloor k \cdot \nicefrac{|S|}{n} \rfloor$.
\smallskip

It is straightforward to observe that Subgraph-Entitlement implies Clique-Entitlement, and, 
consequently, Subgraph-Entitlement is violated by \toprank and \topkatz.  
Before turning to \mesrank and \meskatz,
we take one more step beyond Subgraph-Entitlement, aiming to relax the requirement of strong connectivity among the vertices in $G[S]$. A prominent guarantee in social choice for fairness considerations is the Proportional Justified Representation (PJR) axiom~\citep{sanchez2026proportional}. It ensures that any group of voters comprising at least an $\ell \cdot \nicefrac{n}{k}$ fraction of the population, who all agree on at least $\ell$ common candidates, are collectively guaranteed at least $\ell$ representatives. The strengthening of Subgraph-Entitlement that we define below closely resembles PJR
---this formulation is also conceptually noteworthy, as it repurposes a standard fairness notion from social choice to another domain.

\smallskip
\textbf{Group-Entitlement:}
A rule $\mathcal{R}$ satisfies Group-Entitlement if for every graph $G=(V,E)$, every subset of nodes $S \subseteq V$, and every $\ell \in \mathbb{N}$ such that $|S|\geq \ell \cdot \nicefrac{n}{k}$ and $|\bigcap_{u \in S}\Succ(u)|\geq \ell$, it holds that
$|\Succ(S) \cap \mathcal{R}(G,k)| \ge \ell.$
\smallskip

\mesrank \textit{and} \meskatz \textit{satisfy Group-Entitlement, and, as a byproduct, also Clique- and Subgraph-Entitlement.}
This provides a solid proportionality guarantee: any sufficiently large group of nodes with substantial overlap in their successors is entitled to proportional representation in the outcome.

We highlight that our positive axiomatic results are not specific to PageRank and Katz; they hold for any rule that combines the Method of Equal Shares with any measure of nodes' importance, as long as a basic consistency condition is met: the utility of a node $u$ from a node $v,$ as per the function $\mu$, is non-zero if and only if 
there exists a walk from $u$ to $v$, i.e., 
$u$ is connected to $v$, at least indirectly.

\section{Experimental Evaluation}
\label{sec:experiments}

In this section, we analyze the discussed rules
empirically, on both real-life and synthetic data.
Additionally, we consider \bosrank and \boskatz, which are
defined similarly to \mesrank and \meskatz,
but using the Method of Equal Shares with Bounded Overspending (BOS)~\citep{papasotiropoulos2024method}, i.e.,
the fine-tuned variant of MES which is argued to better handle data with high variance in candidate utilities,
a common characteristic of our model.
We fix $\alpha=0.85$ for PageRank, as used, e.g., by \citet{brin1998anatomy}
and $\alpha=0.85/\lambda$ for Katz-based rules.

We first note that the outcome of \mesrank (and similarly \meskatz) can be computed in polynomial time.
This is because PageRank and, in particular, the utility function from \Cref{eq:pagerank_evaluations}, can be computed in polynomial time~\citep{langville2006google}, specifically in $O(mn)$ by the power iteration method---while also being parallelizable.
The same applies to the Method of Equal Shares, which runs in $O(kn^2)$.
Therefore, we begin by analyzing the running times of our rules
and compare them to the running time of 
local search group-closeness centrality~\citep{AngBecGilDerEtal-2021-GroupCloseness},
which aims at selecting a group of nodes 
such that the average distance from every node in the graph
to the closest one in the group is minimized.
To this end, we compute the outcomes of all considered rules
on randomly generated directed graphs where
for each pair of nodes $i, j$ we choose
whether there will be edge $(i,j)$ independently
with probablity $p$ 
(which is a standard adaptation of the Erdős-Rényi model~\citep{ErdRen-1959-RandomGraphs,Gil-1959-RandomGraphs} to directed graphs).
We select $p$ such that the average out-degree of each node is $5$.
First, we ccompute the outcomes for constant number of selected nodes, $10$,
but we vary the size of the graph,
and then we fix the size of the graph on $500$ nodes, but vary the number of selected nodes.
The results show that all \textsc{Mes} and \textsc{Bos} methods scale very well
and run not much slower than the simple \textsc{Top} approach.
In particular, all rules run much faster than the local search group-closeness.
Average running times are shown at \Cref{fig:runtime:plots:appendix}.\footnote{
	The computations have been done in Python
	using a MacBook Air laptop with an Apple M3 processor and 16 GB of RAM.}
\begin{figure*}[t]
	\centering
	\pgfdeclareplotmark{mystar}{
    \node[star,star point ratio=2.25,minimum size=6pt,
          inner sep=0pt,draw=none,solid,fill=Plum] {};
}
\begin{tikzpicture}
    \draw [draw=white] (-0.9,3) rectangle (0,3.5);
    \draw [draw=black] (0,3) rectangle (14.7,3.5);

    \node (top_s) at (0.2, 3.25) {};
    \node (top_e) at (1.3, 3.25) {\footnotesize \textsc{Top}};
    \node [circle, minimum size=0.2cm, fill=BrickRed, inner sep=0pt] (_) at (0.68, 3.25) {};
    \node (mes_s) at (1.7, 3.25) {};
    \node (mes_e) at (2.8, 3.25) {\footnotesize \textsc{Mes}};
    \node [diamond, minimum size=0.2cm, fill=NavyBlue, inner sep=0pt] (_) at (2.15, 3.25) {};
    \node (bos_s) at (3.2, 3.25) {};
    \node (bos_e) at (4.3, 3.25) {\footnotesize \textsc{Bos}};
    \node[
        regular polygon,
        regular polygon sides=4,
        minimum size=6pt,
        inner sep=0pt,
        draw=none,
        fill=ProcessBlue
        ] (_) at (3.65, 3.25) {};
    \node (clo_s) at (4.6, 3.25) {};
    \node (clo_e) at (6, 3.25) {\footnotesize \textsc{Closeness}};
    \node[star,star point ratio=2.25,minimum size=6pt,
        inner sep=0pt,draw=none,solid,fill=Plum]
        (_) at (5.05, 3.258) {};
    \node (tim_s) at (6.8, 3.25) {};

    \path[thick]
    (top_s) edge[draw=BrickRed] (top_e)
    (mes_s) edge[draw=NavyBlue] (mes_e)
    (bos_s) edge[draw=ProcessBlue] (bos_e)
    (clo_s) edge[draw=Plum] (clo_e)
    ;
\end{tikzpicture}
\begin{tikzpicture}
    \begin{axis}[
        name=runtime_plot_pr_n,
        width=5cm,
        height=5.5cm,
        title={PageRank rules},
        title style={yshift=-6pt},
        xtick={250, 500, 750, 1000},
        xlabel={no. nodes},
        ylabel={avg. running time (s.)},
        xlabel style={align=center,yshift=8pt},   
        ylabel style={align=center,yshift=-16pt},   
        ytick={0, 2, 4, 6, 8, 10},
        ymin=-0.5, ymax=10.5,
        legend pos=north east,
        legend cell align={left},
        legend style={nodes={scale=0.8, transform shape}},
        tick label style={font=\scriptsize},
        label style={font=\small},
        legend style={font=\small}
    ]
    
    \addplot[color=BrickRed,mark=*,thick,error bars/.cd,y dir=both,y explicit]
    table[x expr=\thisrowno{0}-0.02, y index=1, y error index=2] {data/runtime_rnd_graph_n_pr.dat};

    \addplot[color=NavyBlue,mark=diamond*,thick,error bars/.cd,y dir=both,y explicit]
    table[x expr=\thisrowno{0}-0.01, y index=3, y error index=4] {data/runtime_rnd_graph_n_pr.dat};

    \addplot[color=ProcessBlue,mark=square*,thick,error bars/.cd,y dir=both,y explicit]
    table[x expr=\thisrowno{0}, y index=5, y error index=6] {data/runtime_rnd_graph_n_pr.dat};

    \addplot[color=Plum,mark=mystar,thick,error bars/.cd,y dir=both,y explicit]
    table[x expr=\thisrowno{0}+0.01, y index=7, y error index=8] {data/runtime_rnd_graph_n_dist.dat};
    
    \end{axis}
    \begin{axis}[
        name=runtime_plot_katz_n,
        at={(runtime_plot_pr_n.south east)},
        title ={Katz Rules},
        title style={yshift=-3pt},
        width=5cm,
        height=5.5cm,
        ytick={0, 2, 4, 6, 8, 10},
        ymin=-0.5, ymax=10.5,
        yticklabels={{}},
        xtick={250, 500, 750, 1000},
        xlabel={no. nodes},
        tick label style={font=\scriptsize},
        label style={font=\small},
        xlabel style={align=center,yshift=8pt}
    ]
    \addplot[color=BrickRed,mark=*,thick,error bars/.cd,y dir=both,y explicit]
    table[x expr=\thisrowno{0}-0.02, y index=1, y error index=2] {data/runtime_rnd_graph_n_katz.dat};

    \addplot[color=NavyBlue,mark=diamond*,thick,error bars/.cd,y dir=both,y explicit]
    table[x expr=\thisrowno{0}-0.01, y index=3, y error index=4] {data/runtime_rnd_graph_n_katz.dat};

    \addplot[color=ProcessBlue,mark=square*,thick,error bars/.cd,y dir=both,y explicit]
    table[x expr=\thisrowno{0}, y index=5, y error index=6] {data/runtime_rnd_graph_n_katz.dat};

    \addplot[color=Plum,mark=mystar,thick,error bars/.cd,y dir=both,y explicit]
    table[x expr=\thisrowno{0}+0.01, y index=7, y error index=8] {data/runtime_rnd_graph_n_dist.dat};

    \end{axis}
\end{tikzpicture}
\begin{tikzpicture}
    \begin{axis}[
        name=runtime_plot_pr_k,
        width=5cm,
        height=5.5cm,
        title={PageRank rules},
        title style={yshift=-6pt},
        xlabel={\small no. selected nodes},
        ylabel={\small avg. running time (s.)},
        xlabel style={align=center,yshift=8pt},   
        ylabel style={align=center,yshift=-16pt},   
        xtick={10, 15, 20, 25},
        ytick={0, 2, 4, 6, 8, 10},
        ymin=-0.5, ymax=10.5,
        tick label style={font=\scriptsize},
    ]
    \addplot[color=BrickRed,mark=*,thick,error bars/.cd,y dir=both,y explicit]
    table[x expr=\thisrowno{0}-0.02, y index=1, y error index=2] {data/runtime_rnd_graph_k_pr.dat};

    \addplot[color=NavyBlue,mark=diamond*,thick,error bars/.cd,y dir=both,y explicit]
    table[x expr=\thisrowno{0}-0.01, y index=3, y error index=4] {data/runtime_rnd_graph_k_pr.dat};

    \addplot[color=ProcessBlue,mark=square*,thick,error bars/.cd,y dir=both,y explicit]
    table[x expr=\thisrowno{0}, y index=5, y error index=6] {data/runtime_rnd_graph_k_pr.dat};

    \addplot[color=Plum,mark=mystar,thick,error bars/.cd,y dir=both,y explicit]
    table[x expr=\thisrowno{0}+0.01, y index=7, y error index=8] {data/runtime_rnd_graph_k_dist.dat};
    
    \end{axis}
    \begin{axis}[
        name=runtime_plot_katz_k,
        at={(runtime_plot_pr_k.south east)},
        title={Katz rules},
        title style={yshift=-3pt},
        width=5cm,
        height=5.5cm,
        ytick=\empty,
        xtick={10, 15, 20, 25},
        ytick={0, 2, 4, 6, 8, 10},
        ymin=-0.5, ymax=10.5,
        yticklabels = {{}},
        tick label style={font=\scriptsize},
        label style={font=\small},
        xlabel={{\small no. selected nodes}},
        xlabel style={align=center,yshift=8pt}
    ]
    \addplot[color=BrickRed,mark=*,thick,error bars/.cd,y dir=both,y explicit]
    table[x expr=\thisrowno{0}-0.02, y index=1, y error index=2] {data/runtime_rnd_graph_k_katz.dat};

    \addplot[color=NavyBlue,mark=diamond*,thick,error bars/.cd,y dir=both,y explicit]
    table[x expr=\thisrowno{0}-0.01, y index=3, y error index=4] {data/runtime_rnd_graph_k_katz.dat};

    \addplot[color=ProcessBlue,mark=square*,thick,error bars/.cd,y dir=both,y explicit]
    table[x expr=\thisrowno{0}, y index=5, y error index=6] {data/runtime_rnd_graph_k_katz.dat};

    \addplot[color=Plum,mark=mystar,thick,error bars/.cd,y dir=both,y explicit]
    table[x expr=\thisrowno{0}+0.01, y index=7, y error index=8] {data/runtime_rnd_graph_k_dist.dat};

    \end{axis}



\end{tikzpicture}
	\caption{		
		The average time needed to compute the outcomes of our rules on random graphs
		depending on the number of nodes (first two plots)
		or the number of nodes in the selection (last two plots).
For group-closeness, we omit extreme values from the plot to preserve the visibility of differences between the rest of the rules.}
	\label{fig:runtime:plots:appendix}
\end{figure*}

In what follows, we present the main experimental evaluation of our study. In particular,
first, we analyze the proportionality aspects of our rules
based on two network datasets often used as benchmarks for community detection algorithms~\citep{jin2021survey}.
Then we extend our study by analyzing larger real-world data as well as synthetic datasets.
Whenever undirected graphs are used, we convert them into directed graphs
by replacing each undirected edge with two directed edges in opposite directions.
Our experiments demonstrate that 
the proposed methods can proportionally represent different communities of the network and that
relying on the network structure is enough to enable fair selection that reflects the underlying structure even with respect to groups that are unknown or not explicitly defined.

\subsection{College Football Network}\label{sec:college-football-network}
The first dataset~\citep{GirNew-2002-CommunityDetection} is a graph of 115 nodes representing U.S. college football teams.
Each undirected edge is a game played in Division IA during the 2000 Fall season. 
The teams are split into 11 conferences and a group of independents;
64\% of games occur within the conferences.
Refer to \Cref{fig:football} (Left) for an illustration, in which each group of nodes represents a conference or a set of independent teams.
It also shows the outcomes of \topkatz and \meskatz for $k=8$. 
There are two conferences from which \topkatz selects three nodes,
and hence it covers only four conferences in total.
In contrast,
\meskatz distributes its selection more evenly across conferences, selecting at most one team per conference.
The conclusions for \toprank and \mesrank are analogous. 
To generalize this analysis for other values of $k$,
we plot the maximum number of teams selected from a single conference for each rule
and for $k \in \{1,\dots,50\}$ (see \Cref{fig:football} (Right)).
We also include \boskatz and \bosrank there.
The result shows that as we increase $k$,
the difference between \topkatz/\toprank and the other rules becomes even more pronounced.

\begin{figure}[t]
	\centering
		\scalebox{0.9}{\input{figures/football_network.tex}}
	\hspace{0.7cm}
		\scalebox{0.9}{\begin{tikzpicture}
    \draw [draw=black] (0,3) rectangle (7.83,3.5);
    \node (top_s) at (0.5,3.25) {};
    \node (top_e) at (1.6,3.25) {\textsc{Top}};
    \node (mes_s) at (2.5,3.25) {};
    \node (mes_e) at (3.6,3.25) {\textsc{Mes}};
    \node (bos_s) at (4.5,3.25) {};
    \node (bos_e) at (5.6,3.25) {\textsc{Bos}};

    \path[very thick]
    (top_s) edge[draw=BrickRed!20] (top_e)
    (top_s) edge[draw=BrickRed, dashdotted] (top_e)
    (mes_s) edge[draw=NavyBlue] (mes_e)
    (bos_s) edge[draw=ProcessBlue!20] (bos_e)
    (bos_s) edge[draw=ProcessBlue, dotted] (bos_e)
    ;

    \begin{axis}[
        name=pr_plot,
        width=5.5cm,
        height=4.5cm,
        xlabel={committee size\\(PageRank rules)},
        ylabel={max. from 1 conf.},
        xlabel style={align=center,yshift=5pt},   
        ylabel style={yshift=-16pt},   
        xtick={1,10,20,30,40,50},
        xticklabels={{1}, {10}, {20}, {30}, {40}, {50}},
        ymin=0.5, ymax=12,
        ytick={1, 3, 5, 7, 9, 11}, 
        legend pos=north east,
        legend cell align={left},
        legend style={nodes={scale=0.8, transform shape}},
        tick label style={font=\scriptsize},
        label style={font=\small},
        legend style={font=\small}
    ]
    
    \addplot[color=BrickRed!20,very thick]
    table[x, y expr=\thisrowno{1}+0.4] {data/pr_rules_max_conf.dat};
    \addplot[color=BrickRed,very thick,dashdotted]
    table[x, y expr=\thisrowno{1}+0.4] {data/pr_rules_max_conf.dat};



    \addplot[color=NavyBlue,very thick]
    table[x, y  expr=\thisrowno{3}+0.0] {data/pr_rules_max_conf.dat};

    \addplot[color=ProcessBlue!20,very thick]
    table[x, y  expr=\thisrowno{2}-0.2] {data/pr_rules_max_conf.dat};
    \addplot[color=ProcessBlue,very thick,dotted]
    table[x, y  expr=\thisrowno{2}-0.2] {data/pr_rules_max_conf.dat};
    
    \end{axis}
    \begin{axis}[
        name=katz_plot,
        at={(pr_plot.south east)},
        width=5.5cm,
        height=4.5cm,
        ytick=\empty,
        xtick={1,10,20,30,40,50},
        xticklabels={{1}, {10}, {20}, {30}, {40}, {50}},
        tick label style={font=\scriptsize},
        label style={font=\small},
        xlabel={committee size\\(Katz rules)},
        xlabel style={align=center,yshift=5pt},
        ymin=0.5, ymax=12,
    ]
    \addplot[color=BrickRed!20,very thick]
    table[x, y expr=\thisrowno{1}+0.4] {data/katz_rules_max_conf.dat};
    \addplot[color=BrickRed,very thick,dashdotted]
    table[x, y expr=\thisrowno{1}+0.4] {data/katz_rules_max_conf.dat};



    \addplot[color=NavyBlue,very thick]
    table[x, y  expr=\thisrowno{3}+0.0] {data/katz_rules_max_conf.dat};

    \addplot[color=ProcessBlue!20,very thick]
    table[x, y  expr=\thisrowno{2}-0.2] {data/katz_rules_max_conf.dat};
    \addplot[color=ProcessBlue,very thick,dotted]
    table[x, y  expr=\thisrowno{2}-0.2] {data/katz_rules_max_conf.dat};
    \end{axis}
\end{tikzpicture}}
	\caption{The College Football Network and the outcomes of 
		\topkatz (\textcolor{winered}{red} double lines) and \meskatz (\textcolor{NavyBlue}{blue} pattern) for $k=8$ (Left) and 
		the maximum number of nodes selected from a single conference for a given committee size (Right).}
	\label{fig:football}
\end{figure}

\subsection{Political Blogs Network}
\label{sec:polblogs}

The second network we analyze~\citep{AdaGla-2005-PolBlogs} consists of 1,490 nodes representing political blogs that were active during the 2004 U.S. presidential election. 
An edge from blog A to blog B indicates a front-page link from A to B. 
Blogs are labeled as \emph{liberal} or \emph{conservative}, with a roughly balanced composition: 758 liberal and 732 conservative.
To assess how our rules perform on unbalanced data,
we delete each node of a given fixed label with probability $p$
and evaluate how well the label distribution in the outcomes
reflects that in the modified graph.
We also check the effectiveness of the selection
using three metrics:
summed PageRanks of the selected nodes;
summed Katz centralities; and
an average spread of influence under the \emph{independent cascades} model~\citep{kempe2003maximizing} with the selected nodes as seed
and uniform transmission probability of $2\%$
(a low value is used since the network is very dense). As for the latter, recall that, roughly speaking, in the independent cascades model
	with uniform transmission probability $\tau$,
	in each round, for each node $v$ that was infected in a previous round,
	we look at each of its incoming edges
	and infect the start of this edge with probability $\tau$
	(usually the infection progresses with outgoing edges,
	but for our network the reverse approach is more natural).
	Seeds are infected in round $0$.

As baselines we include a random selection of nodes, the
local search group-closeness centrality~\citep{AngBecGilDerEtal-2021-GroupCloseness}
that aims at diversity,
and a standard influence maximization algorithm \tim~\citep{tang2014tim_plus}
(we also considered other influence maximization algorithms like \textsc{Celf}++~\citep{goyal2011celf},
but their running times were prohibitive for our experiment).
For each $p \in \{0.1, 0.3, 0.5, 0.7, 0.9\}$, we generated 100 graphs:
50 graphs with liberal nodes removed, and 50 with conservative.  
Figure~\ref{fig:polblogs:plots_PR} shows the results of our analysis for $k=10$
(results for $k=20$ are similar, see the Appendix).

\begin{figure*}[t]
	\input{figures/polblog_plots_main_body_full.tex}
	\captionof{figure}{
		Results of our experiments on Political Blogs Network for $k=10$.
		Each column presents the values of a different statistic
		depending on the probability of removing blogs from one side ($x$-axis).
		In the first column, we show the average
		fraction of minority-labeled nodes in the outputs of the rules
		(in this column we also present the fraction of minority-labeled nodes
		among all nodes in the network---grey dashed line).
		In the second and third columns, we plot the
		average summed values of PageRank and Katz centrality, respectively,
		among the selected nodes.
		In the fourth column,
		we present the average number of infected nodes
		in the independent cascades model,
		when the seed is an output of a rule.
		Top row shows the values for \toprank, \mesrank, and \bosrank,
		while the bottom one for \topkatz, \meskatz, and \boskatz.
		The values for the remaining methods are identical in both rows.
		Vertical bars denote 95\% confidence intervals.
	}
\label{fig:polblogs:plots_PR}
\end{figure*}

For PageRank-based rules, the proportions of nodes with the
two labels in the input graphs are relatively well reflected
in the outcomes of our rules for small values of $p$.
For larger values of $p$,
\random selection performs clearly better than other rules,
while \bosrank is the second best.
\toprank, \mesrank, and \tim significantly underrepresent the minority for larger values of $p$,
while
group-closeness consistently overrepresents the minority.

For the Katz-based rules, the differences are stark.
\boskatz closely follows the proportions in the input graphs
(outperforming \bosrank in this aspect),
\meskatz performs a bit worse,
while \topkatz completely excludes minority nodes even for moderate values of $p$.

Next, we look at the total PageRank and Katz centrality of the selected nodes,
which serve as metrics for selection of the most influential nodes.
The pattern is the same for both metrics:
\top and \mes approaches perform similarly well,
while \bos based rules exhibits a minimally worse performance.
All of our baselines are significantly far behind,
with \tim performing reasonably well for larger values of $p$.

For the average spread under the independent cascades model
we see a similar pattern to that for PageRank and Katz centrality.
This is quite surprising given that \tim
is an algorithm designed for that purpose
(and it was given the exact parameters of the model,
including transmission probability,
while the remaining methods were blind to it).
However, imperfect performance of this algorithm
in certain settings has been noted before~\citep{arora2017debunking}.

Only our election-based methods are consistently performing well
across all of the examined criteria.
The Katz-based rules perform well even when evaluated
under total achieved PageRank centrality,
showing the robustness of this approach.
Thus, \boskatz gives almost perfectly proportional representation
of the groups in the network, while performing very well under all three efficiency metrics.

\subsection{Further Real-World Data}

We analyze two additional networks, both relatively large and with nodes split into categories.
The first is a network of Facebook pages, where nodes represent official pages and edges denote mutual likes~\citep{rozemberczki2019multiscale}. It includes pages from four categories: politicians, governmental organizations, TV shows, and companies.
The second is a CiteSeer network, where nodes represent articles labeled with academic categories~\citep{network_repository}.
For both datasets, we report the $\ell_1$ distance between the vector of frequencies of different labels in the outputs of our rules and the corresponding vector for the entire network. 
We report results for $k=10, 20$ and $50$ in \Cref{table:freqs-in-nets_full}.
For the rules based on Katz, the pattern is clear: \boskatz consistently achieves the minimum distances among Katz-based rules, while \meskatz always performs significantly better than \topkatz.
For PageRank, \bosrank is almost always performing better than \mesrank and \toprank; with the exception of the Facebook network and $k=10$.
There, the distance of \bosrank is larger than that of \toprank and \mesrank.
Two factors might have resulted in that.
First, for $k=10$, selecting one node instead of another may change the distance significantly,
hence there is a high degree of randomness in the outcomes.
Second, $0.79$ is not a bad outcome,
but the other rules, including \toprank, seem to perform quite well in this particular case.
This is due to the fact that PageRank in itself gives somewhat proportional outcomes,
which we discuss in details in \Cref{sec:concl}.

\begin{table*}[t]
	\centering
	\bigskip
	\resizebox{0.75\columnwidth}{!}{%
	\setlength{\tabcolsep}{6pt}
	\begin{tabular}{cc | ccc | ccc | ccc}
		&   & \multicolumn{3}{c}{$k = 10$} & \multicolumn{3}{|c}{$k = 20$} & \multicolumn{3}{|c}{$k = 50$} \\
		\toprule
		network & centrality  & \textsc{Top} & \textsc{Mes} & \textsc{Bos} & \textsc{Top} & \textsc{Mes} & \textsc{Bos} &  \textsc{Top}  & \textsc{Mes} & \textsc{Bos} \\
		\midrule
		\multirow{2}{*}{Facebook} & PageRank & $0.68$ &  \cellcolor{mossgreen} $0.41$ & $0.79$ & $0.58$ & $0.49$ & \cellcolor{mossgreen} $0.39$ & $0.51$ & $0.42$ &  \cellcolor{mossgreen} $0.27$ \\
		& Katz & $1.39$ & $0.99$ & \cellcolor{mossgreen} $0.59$ & $1.39$ & $0.99$ &\cellcolor{mossgreen}  $0.49$ & $1.35$ & $0.59$ & \cellcolor{mossgreen}$0.39$ \\
		\multirow{2}{*}{CiteSeer} & PageRank & $0.83$ & $1.03$ & \cellcolor{mossgreen} $0.44$  & $0.53$ & $0.73$ &\cellcolor{mossgreen}  $0.42$  & $0.35$ & $0.55$ & \cellcolor{mossgreen}$0.22$ \\
		& Katz  &  $1.59$ & $1.23$ & \cellcolor{mossgreen} $0.41$ & $1.59$ & $0.83$ &\cellcolor{mossgreen}  $0.28$  & $1.23$ & $0.67$ &\cellcolor{mossgreen}  $0.17$ \\
		\bottomrule
	\end{tabular}}
	\caption{The $\ell_1$ distance between the vector of frequencies of different node labels
		in the outputs of our rules and the corresponding vector of frequencies for the entire network. The smallest distance among rules is highlighted.}
	\label{table:freqs-in-nets_full}
\end{table*}

\subsection{Euclidean Data}

Our final set of experiments is inspired by applications in social choice. As already highlighted, our methods can be applied to elect a committee among a group of voters who vote amongst themselves. In the social choice literature, voters and candidates are often modeled as points in a two-dimensional Euclidean space, typically representing an ideological spectrum~\citep{elk-fal-las-sko-sli-tal:c:2d-multiwinner}. Similarly, we assume that the nodes (representing both voters and candidates) correspond to points. These are sampled from a specific distribution, which we describe later. Edges between nodes are introduced based on the distances between them, using one of four strategies.
The first two strategies align with conventional assumptions in social choice, where voters are more likely to prefer candidates closer to them in the ideological space:
Under \textbf{E-radius}, each node connects (with a fixed probability) to those within a specified radius. 
Under \textbf{E-appr}, each node connects to a fixed number of its closest neighbors. To introduce some noise, we assume that each neighbor can be omitted with a fixed probability.

We also propose a novel Euclidean model, which is well aligned with the idea inspired by liquid democracy: the voters tend to vote for their close friends (i.e., close points), yet in their preferences, exhibit a bias toward candidates with higher competence.
Each candidate is assigned an objective value representing their competence ($y$-coordinate). A node located at point $(x, y)$ connects to the nodes closest to $(x, y + b)$, where $b$ is a constant representing the competence bias. Variants of this model are denoted as \textbf{B-radius} and \textbf{B-appr}, depending on whether the voters approve candidates within a certain radius of acceptability or a certain number of candidates. 

For each of these models, the points are drawn from two Gaussian distributions, with the points divided between the two groups in a $1:3$ or $2:3$ ratio. 
From each setting, we sample 1000 instances, each consists of $n = 1000$ points. For each instance, we construct a graph  with $n$ nodes and select $k = 10$ of them.
We identify the points corresponding to the selected nodes and mark them (in green). The points from the 1000 experiments are combined into a single plot, forming a histogram that represents the distribution of selected points.\looseness-1

\begin{figure*}[!t]
	\centering
	\setlength{\tabcolsep}{5.5pt}
	\scalebox{0.9}{
		\begin{tabular}{cccccccc}
			\multicolumn{1}{c}{E-radius-$\nicefrac{1}{3}$} &\multicolumn{1}{c}{E-radius-$\nicefrac{2}{3}$} &\multicolumn{1}{c}{E-appr-$\nicefrac{1}{3}$} &\multicolumn{1}{c}{E-appr-$\nicefrac{2}{3}$} &\multicolumn{1}{c}{B-radius-$\nicefrac{1}{3}$} & \multicolumn{1}{c}{B-radius-$\nicefrac{2}{3}$} &\multicolumn{1}{c}{B-appr-$\nicefrac{1}{3}$} &\multicolumn{1}{c}{B-appr-$\nicefrac{2}{3}$}  \\
			\multicolumn{8}{c}{}\\ 
			\subcaptionbox*{0.72 : 0.28}{\includegraphics[width=0.1\textwidth]{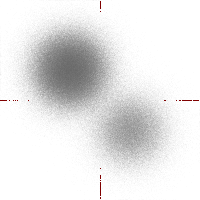}} &
			\subcaptionbox*{0.59 : 0.41}{\includegraphics[width=0.1\textwidth]{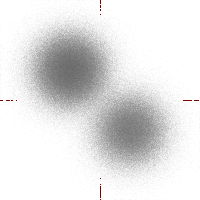}} &
			\subcaptionbox*{0.72 : 0.28}{\includegraphics[width=0.1\textwidth]{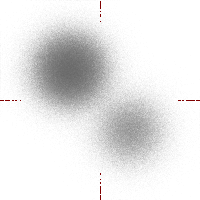}} &
			\subcaptionbox*{0.59 : 0.41}{\includegraphics[width=0.1\textwidth]{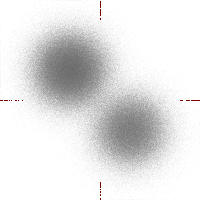}} &
			\subcaptionbox*{0.72 : 0.28}{\includegraphics[width=0.1\textwidth]{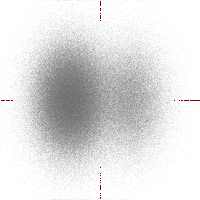}} &
			\subcaptionbox*{0.59 : 0.41}{\includegraphics[width=0.1\textwidth]{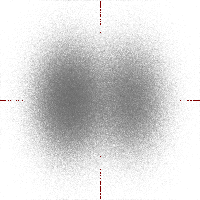}} &
			\subcaptionbox*{0.72 : 0.28}{\includegraphics[width=0.1\textwidth]{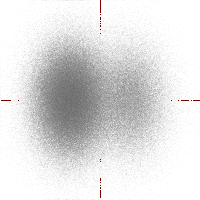}} &
			\subcaptionbox*{0.59 : 0.41}{\includegraphics[width=0.1\textwidth]{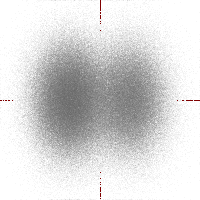}} \\
			\multicolumn{8}{c}{}\\
			\multicolumn{8}{c}{\toprank}\\ 
			\subcaptionbox*{0.41 : 0.59}{\includegraphics[width=0.1\textwidth]{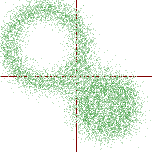}} &
			\subcaptionbox*{0.45 : 0.55}{\includegraphics[width=0.1\textwidth]{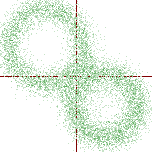}} &
			\subcaptionbox*{0.59 : 0.41}{\includegraphics[width=0.1\textwidth]{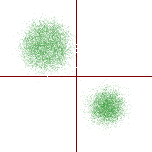}} &
			\subcaptionbox*{0.52 : 0.48}{\includegraphics[width=0.1\textwidth]{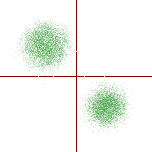}} &
			\subcaptionbox*{0.65 : 0.35}{\includegraphics[width=0.1\textwidth]{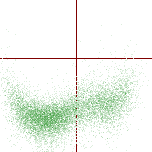}} &
			\subcaptionbox*{0.56 : 0.44}{\includegraphics[width=0.1\textwidth]{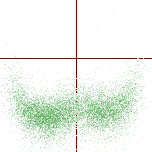}} &
			\subcaptionbox*{0.99 : 0.01}{\includegraphics[width=0.1\textwidth]{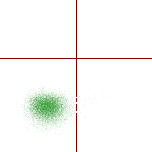}} &
			\subcaptionbox*{0.81 : 0.19}{\includegraphics[width=0.1\textwidth]{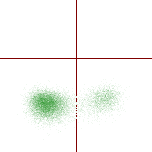}} \\
			\multicolumn{8}{c}{}\\
			\multicolumn{8}{c}{\mesrank}\\ 
			\subcaptionbox*{0.83 : 0.17}{\includegraphics[width=0.1\textwidth]{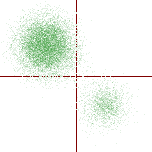}} &
			\subcaptionbox*{0.61 : 0.39}{\includegraphics[width=0.1\textwidth]{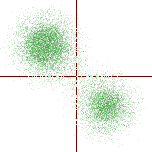}} &
			\subcaptionbox*{0.8 : 0.2}{\includegraphics[width=0.1\textwidth]{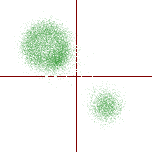}} &
			\subcaptionbox*{0.61 : 0.39}{\includegraphics[width=0.1\textwidth]{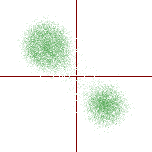}} &
			\subcaptionbox*{0.83 : 0.17}{\includegraphics[width=0.1\textwidth]{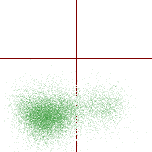}} &
			\subcaptionbox*{0.63 : 0.37}{\includegraphics[width=0.1\textwidth]{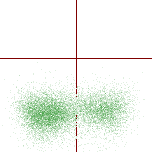}} &
			\subcaptionbox*{0.88 : 0.12}{\includegraphics[width=0.1\textwidth]{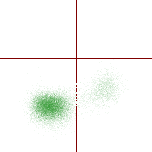}} &
			\subcaptionbox*{0.67 : 0.33}{\includegraphics[width=0.1\textwidth]{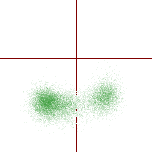}} \\
			\multicolumn{8}{c}{}\\
			\multicolumn{8}{c}{\bosrank}\\ 
			\subcaptionbox*{0.68 : 0.32}{\includegraphics[width=0.1\textwidth]{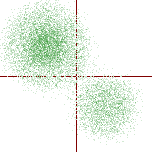}} &
			\subcaptionbox*{0.57 : 0.43}{\includegraphics[width=0.1\textwidth]{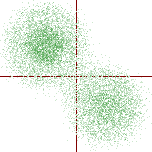}} &
			\subcaptionbox*{0.75 : 0.25}{\includegraphics[width=0.1\textwidth]{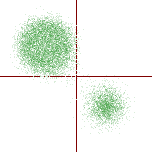}} &
			\subcaptionbox*{0.6 : 0.4}{\includegraphics[width=0.1\textwidth]{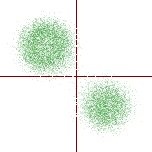}} &
			\subcaptionbox*{0.68 : 0.32}{\includegraphics[width=0.1\textwidth]{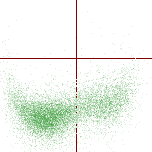}} &
			\subcaptionbox*{0.57 : 0.43}{\includegraphics[width=0.1\textwidth]{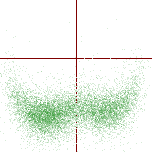}} &
			\subcaptionbox*{0.83 : 0.17}{\includegraphics[width=0.1\textwidth]{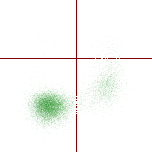}} &
			\subcaptionbox*{0.63 : 0.37}{\includegraphics[width=0.1\textwidth]{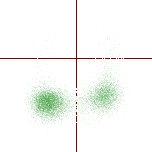}} 
	\end{tabular}}
	\caption{Histograms generated by our PageRank-based methods for Euclidean graphs. The distributions (first row) are presented at their original scale, while the histograms are displayed at 130\% zoom for improved clarity.}
	\label{fig:gridpr}
\end{figure*}

\begin{figure*}[t]
	\centering
	\setlength{\tabcolsep}{5.5pt}		\scalebox{0.9}{
		\begin{tabular}{cccccccc}
			\multicolumn{1}{c}{E-radius-$\nicefrac{1}{3}$} &\multicolumn{1}{c}{E-radius-$\nicefrac{2}{3}$} &\multicolumn{1}{c}{E-appr-$\nicefrac{1}{3}$} &\multicolumn{1}{c}{E-appr-$\nicefrac{2}{3}$} &\multicolumn{1}{c}{B-radius-$\nicefrac{1}{3}$} & \multicolumn{1}{c}{B-radius-$\nicefrac{2}{3}$} &\multicolumn{1}{c}{B-appr-$\nicefrac{1}{3}$} &\multicolumn{1}{c}{B-appr-$\nicefrac{2}{3}$}  \\
			\multicolumn{8}{c}{}\\ 
			\subcaptionbox*{0.72 : 0.28}{\includegraphics[width=0.1\textwidth]{n_1000_k_10_alpha_0_85/setup8/pagerank/points.png}} &
			\subcaptionbox*{0.59 : 0.41}{\includegraphics[width=0.1\textwidth]{n_1000_k_10_alpha_0_85/setup7/pagerank/points.png}} &
			\subcaptionbox*{0.72 : 0.28}{\includegraphics[width=0.1\textwidth]{n_1000_k_10_alpha_0_85/setup8a/pagerank/points.png}} &
			\subcaptionbox*{0.59 : 0.41}{\includegraphics[width=0.1\textwidth]{n_1000_k_10_alpha_0_85/setup7a/pagerank/points.png}} &
			\subcaptionbox*{0.72 : 0.28}{\includegraphics[width=0.1\textwidth]{n_1000_k_10_alpha_0_85/setup4c/pagerank/points.png}} &
			\subcaptionbox*{0.59 : 0.41}{\includegraphics[width=0.1\textwidth]{n_1000_k_10_alpha_0_85/setup3c/pagerank/points.png}} &
			\subcaptionbox*{0.72 : 0.28}{\includegraphics[width=0.1\textwidth]{n_1000_k_10_alpha_0_85/setup4d/pagerank/points.png}} &
			\subcaptionbox*{0.59 : 0.41}{\includegraphics[width=0.1\textwidth]{n_1000_k_10_alpha_0_85/setup3d/pagerank/points.png}} \\
			\multicolumn{8}{c}{}\\
			\multicolumn{8}{c}{\topkatz}\\ 
			\subcaptionbox*{1.0 : 0.0}{\includegraphics[width=0.1\textwidth]{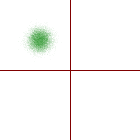}} & 
			\subcaptionbox*{1.0 : 0.0}{\includegraphics[width=0.1\textwidth]{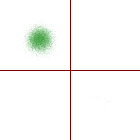}} & 
			\subcaptionbox*{0.6 : 0.4}{\includegraphics[width=0.1\textwidth]{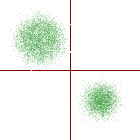}} & 
			\subcaptionbox*{0.53 : 0.47}{\includegraphics[width=0.1\textwidth]{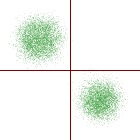}} & 
			\subcaptionbox*{0.78 : 0.22}{\includegraphics[width=0.1\textwidth]{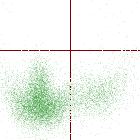}} & 
			\subcaptionbox*{0.71 : 0.29}{\includegraphics[width=0.1\textwidth]{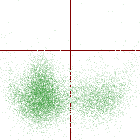}} & 
			\subcaptionbox*{0.99 : 0.01}{\includegraphics[width=0.1\textwidth]{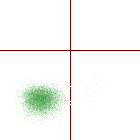}} & 
			\subcaptionbox*{0.81 : 0.19}{\includegraphics[width=0.1\textwidth]{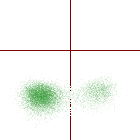}} \\
			\multicolumn{8}{c}{}\\
			\multicolumn{8}{c}{\meskatz}\\ 
			\subcaptionbox*{0.91 : 0.09}{\includegraphics[width=0.1\textwidth]{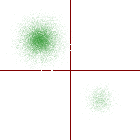}} & 
			\subcaptionbox*{0.68 : 0.32}{\includegraphics[width=0.1\textwidth]{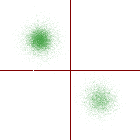}} & 
			\subcaptionbox*{0.8 : 0.2}{\includegraphics[width=0.1\textwidth]{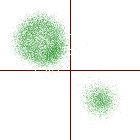}} & 
			\subcaptionbox*{0.61 : 0.39}{\includegraphics[width=0.1\textwidth]{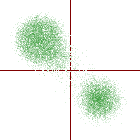}} & 
			\subcaptionbox*{0.83 : 0.17}{\includegraphics[width=0.1\textwidth]{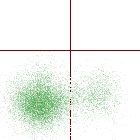}} & 
			\subcaptionbox*{0.63 : 0.37}{\includegraphics[width=0.1\textwidth]{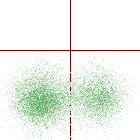}} & 
			\subcaptionbox*{0.88 : 0.12}{\includegraphics[width=0.1\textwidth]{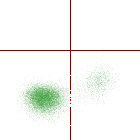}} & 
			\subcaptionbox*{0.67 : 0.33}{\includegraphics[width=0.1\textwidth]{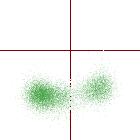}} \\
			\multicolumn{8}{c}{}\\
			\multicolumn{8}{c}{\boskatz}\\ 
			\subcaptionbox*{0.87 : 0.13}{\includegraphics[width=0.1\textwidth]{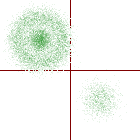}} & 
			\subcaptionbox*{0.61 : 0.39}{\includegraphics[width=0.1\textwidth]{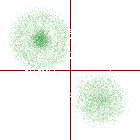}} & 
			\subcaptionbox*{0.75 : 0.25}{\includegraphics[width=0.1\textwidth]{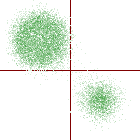}} & 
			\subcaptionbox*{0.6 : 0.4}{\includegraphics[width=0.1\textwidth]{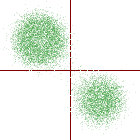}} & 
			\subcaptionbox*{0.72 : 0.28}{\includegraphics[width=0.1\textwidth]{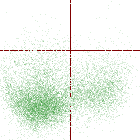}} & 
			\subcaptionbox*{0.59 : 0.41}{\includegraphics[width=0.1\textwidth]{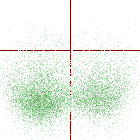}} & 
			\subcaptionbox*{0.83 : 0.17}{\includegraphics[width=0.1\textwidth]{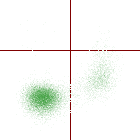}} & 
			\subcaptionbox*{0.64 : 0.36}{\includegraphics[width=0.1\textwidth]{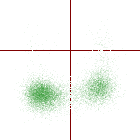}} 
	\end{tabular}}
	\caption{Histograms generated by our Katz-based methods for Euclidean graphs. The distributions (first row) are presented at their original scale, while the histograms are displayed at 150\% zoom for improved clarity.}
	\label{fig:gridkatz}
\end{figure*}

The histograms are presented in \Cref{fig:gridpr} and \Cref{fig:gridkatz}.
The first row shows the distribution of the points, while the subsequent rows illustrate the histograms generated for our rules.
Additionally, below each histogram, we provide the proportion of points selected from each half of the histogram.
Our conclusions are as follows.
First, we observe that \boskatz and \bosrank most accurately reflect
the original proportions of the points in the two Gaussians;
other PageRank rules and \meskatz perform slightly worse,
while \absorbkatz performs significantly worse,
particularly when edges are formed based on radii.
\topkatz often fails to provide any proportionality,
which strongly motivates our proposed methods.
Second, we confirm that our methods recover competence bias from the graph,
selecting nodes corresponding to high competence.
Interestingly, these are not necessarily the nodes with the highest in-degree.

\section{Concluding Discussion}
\label{sec:concl}

We introduced the problem of selecting $k$ influential nodes from a network while ensuring proportional representation of different groups implicitly present in it. 
We proposed a general technique of extending a given centrality measure to a proportional node-selection method.

We observe that all of our proposed solutions enable a significantly more representative selection of nodes compared to the existing \top approach.
The difference is particularly pronounced for Katz centrality, but even for PageRank we see noticeable improvements across different datasets.
This effect is especially evident in Euclidean graphs and in the College Football Network, and reinforced by the theoretical analysis.
While PageRank and Katz centrality are based on a similar idea, they often lead to substantially different outcomes. Below we further discuss the differences between the two approaches, and their suitability depending on the application at hand.

We observe that the straightforward approach of selecting the $k$ nodes with the highest centralities leads to a more proportional representation under PageRank than under Katz centrality. This suggests that PageRank is inherently more proportional.
To some extent, this can be explained by the fact that in PageRank every node distributes its contribution evenly among the centralities of other nodes, leaving no possibility of increasing one's impact. In particular, this means that PageRank significantly limits the influence of even the most influential nodes with higher out-degree.
The strict dampening of influence based on out-degree is often undesirable, making PageRank unsuitable for certain scenarios. For example, if we interpret edges as votes, we previously noted that PageRank extends satisfaction approval voting, while Katz follows the logic of approval voting. In social choice theory, approval voting is often considered the preferable method and is widely used in practice~\citep{av-handbook}. 
This observation is further supported by \Cref{fig:gridpr}, where we can observe that PageRank tends to select more extremist nodes in the surrounding of the two Gaussians. This occurs because these nodes are less ``distracted''---having fewer outgoing edges---making them more likely to be selected. In a way the candidates coming from less popular regions of voters preferences are additionally privileged, an arguably very undesired behaviour in election context. 
Additionally, the axiomatic analysis of PageRank per se suggests that it is unsuitable for certain applications~\citep{WasSki-2023-PageRank}. 
In such applications, our work mitigates the limitations of PageRank by allowing the use of different centrality measures
or machine learning models tailored to the specific application
while still providing a method for selecting representative nodes.

Among the studied rules, those based on the Method of Equal Shares with Bounded Overspending
yield the most representative committees, regardless of whether they are combined with PageRank or Katz centrality. 
The universality of the BOS approach suggests that it is also a preferable method
to be combined with measures beyond the ones studied in this paper.

\section*{Acknowledgements}
Tomasz \Was{} was supported by UK Engineering and Physical Sciences Research Council (EPSRC) under grant EP/X038548/1.
The rest of the authors were partially supported by the European Union (ERC, PRO-DEMOCRATIC, 101076570). Views and opinions expressed are however those of the authors only and do not necessarily reflect those of the European Union or the European Research Council. Neither the European Union nor the granting authority can be held responsible for them.
Oskar Skibski was supported by the National Science Centre under Grant No. 2023/50/E/ST6/00665.
\begin{figure}[h]
	\begin{center}
		\includegraphics*[scale=0.18]{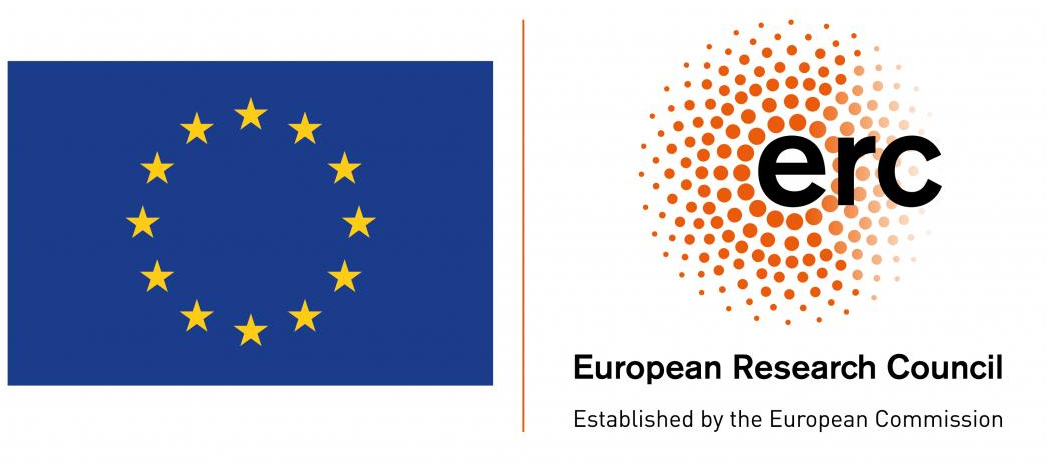}	
	\end{center}
\end{figure}

\bibliography{sample-base}

@String{Computing = "Computing" }

@String{Computer = "{IEEE} Computer" }

@String{Psychometrika = "Psychometrika" }

@String{Springer = "Springer-Verlag" }

@inproceedings{lyu2021centrality,
  title={Centrality with diversity},
  author={L.~Lyu and B.~Fain and K.~Munagala and K.~Wang},
  booktitle={Proceedings of the ACM International Conference on Web Search and Data Mining},
  pages={644--652},
  year={2021}
}

@inproceedings{lin2010improving,
  title={Improving diversity in Web search results re-ranking using absorbing random walks},
  author={G.-L.~Lin and H.~Peng and Q.-L.~Ma and J.~Wei and J.-W.~Qin},
  booktitle={International Conference on Machine Learning and Cybernetics},
  pages={2416--2421},
  year={2010}}

@inproceedings{kempe2003maximizing,
  title={Maximizing the spread of influence through a social network},
  author={D.~Kempe and J.~Kleinberg and E.~Tardos},
  booktitle={Proceedings of the International Conference on Knowledge Discovery and Data Mining},
  pages={137--146},
  year={2003}
}

@inproceedings{network_repository,
  title={The Network Data Repository with Interactive Graph Analytics and Visualization},
  author={R.~A.~Rossi and N.~K.~Ahmed},
  booktitle={Proceedings of the AAAI Conference on Artificial Intelligence},
  year={2015},
  Pages = {4292--4293},
}

@book{av-handbook,
  title = "Handbook on Approval Voting",
  Publisher = "Springer",
  Editor = "J.-F.~Laslier and M.~R.~Sanver",
  year = "2010",
}

@inproceedings{elk-fal-las-sko-sli-tal:c:2d-multiwinner,
  Author = "E.~Elkind and P.~Faliszewski and J.~Laslier and P.~Skowron and A.~Slinko and N.~Talmon",
  Title = "What do multiwinner voting rules do? {An} experiment over the two-dimensional Euclidean domain",
  booktitle={Proceedings of the AAAI Conference on Artificial Intelligence},
  Year = 2017,
  pages = "494--501",
  }

@InProceedings{alon2011sum,
  author    = {N.~Alon and F.~Fischer and A.~Procaccia and M.~Tennenholtz},
  title     = {Sum of us: Strategyproof selection from the selectors},
  booktitle = {Proceedings of the Conference on Theoretical Aspects of Rationality and Knowledge},
  year      = {2011},
  pages     = {101--110},
}

@Article{rozemberczki2019multiscale,
  title = {Multi-scale Attributed Node Embedding},
  author = {B.~Rozemberczki and C.~Allen and R.~Sarkar},
  journal={Journal of Complex Networks},
  year={2021},
  volume={9},
  number={2},
  publisher={Oxford University Press},
  pages = {cnab014}
}

@InProceedings{liquid_survey,
  title = {Modeling Liquid Democracy: A Survey of the (Computational) Social Choice Literature},
  author = {D.~Grossi and A.~Nitsche and G.~Papasotiropoulos},
  booktitle = {Proceedings of the International Joint Conference on Artificial Intelligence},
  year = {2026},
  note = {forthcoming}
}

@inproceedings{brill2018interactive,
  title={Interactive democracy},
  author={M.~Brill},
  booktitle={Proceedings of the International Conference on Autonomous Agents and MultiAgent Systems},
  pages={1183--1187},
  year={2018}
}

@inproceedings{aziz2016strategyproof,
  title={Strategyproof peer selection: Mechanisms, analyses, and experiments},
  author={H.~Aziz and O.~Lev and N.~Mattei and J.~Rosenschein and T.~Walsh},
  booktitle={Proceedings of the AAAI Conference on Artificial Intelligence},
  year={2016},
  pages={390--396},
}

@Article{boldi2011viscous,
  author    = {P.~Boldi and F.~Bonchi and C.~Castillo and S.~Vigna},
  title     = {Viscous democracy for social networks},
  journal   = {Communications of the ACM},
  year      = {2011},
  volume    = {54},
  number    = {6},
  pages     = {129--137},
  publisher = {ACM New York, NY, USA},
}

@InProceedings{AngBecGilDerEtal-2021-GroupCloseness,
  author       = {E.~Angriman and R.~Becker and G.~d'Angelo and H.~Gilbert and A.~van Der Grinten and H.~Meyerhenke},
  title        = {Group-harmonic and group-closeness maximization---approximation and engineering},
  booktitle    = {Proceedings of the Workshop on Algorithm Engineering and Experiments},
  year         = {2021},
  pages        = {154--168}
}

@incollection{FSST-trends,
  author = {P.~Faliszewski and P.~Skowron and A.~Slinko and N.~Talmon},
  booktitle = {Trends in Computational Social Choice},
  title = {Multiwinner Voting: {A} new challenge for social choice theory},
  publisher = "AI Access",
  year = {2017}}

@book{lac-sko:b:approval-survey,
  author       = {M.~Lackner and P.~Skowron},
  title        = {Multi-winner voting with approval preferences},
  series       = {Springer Briefs in Intelligent Systems},
  publisher    = {Springer},
  year         = {2023},
}

@TechReport{PagBriMotWin-1999-PageRank,
  author      = {L.~Page and S.~Brin and R.~Motwani and T.~Winograd},
  title       = {The {P}age{R}ank citation ranking: Bringing order to the web.},
  institution = {Stanford Infolab},
  year        = {1999},
}

@article{brin1998anatomy,
  title={The anatomy of a large-scale hypertextual web search engine},
  author={S.~Brin and L.~Page},
  journal={Computer Networks and ISDN Systems},
  volume={30},
  number={1-7},
  pages={107--117},
  year={1998},
  publisher={Elsevier}
}

@Article{WasSki-2023-PageRank,
  author    = {T.~\Was{} and O.~Skibski},
  title     = {Axiomatic characterization of {P}age{R}ank},
  journal   = {Artificial Intelligence},
  year      = {2023},
  note      = {103900},
  publisher = {Elsevier},
}

@Article{Kat-1953-KatzCentrality,
  author    = {L.~Katz},
  title     = {A new status index derived from sociometric analysis},
  journal   = {Psychometrika},
  year      = {1953},
  volume    = {18},
  number    = {1},
  pages     = {39--43},
  publisher = {Springer},
}

@Article{aziz2017justified,
  author    = {H.~Aziz and M.~Brill and V.~Conitzer and E.~Elkind and R.~Freeman and T.~Walsh},
  title     = {Justified representation in approval-based committee voting},
  journal   = {Social Choice and Welfare},
  year      = {2017},
  volume    = {48},
  number    = {2},
  pages     = {461--485},
  publisher = {Springer},
}

@InProceedings{Peters:Skowron:2020,
  author    = {D.~Peters and P.~Skowron},
  title     = {Proportionality and the limits of welfarism},
  booktitle = {Proceedings of the ACM Conference on Economics and Computation},
  year      = {2020},
  pages     = {793--794},
}

@InProceedings{papasotiropoulos2024method,
  author       = {G.~Papasotiropoulos and Z.~Pishbin and O.~Skibski and P.~Skowron and T.~\Was},
  title        = {Method of equal shares with bounded overspending},
  booktitle    = {Proceedings of the {ACM} Conference on Economics and Computation},
  pages        = {841--868},
  publisher    = {{ACM}},
  year         = {2025}}

@InProceedings{fal-fli-pet-pie-sko-sto-szu-tal:pb-experiments,
  author       = {P.~Faliszewski and J.~Flis and D.~Peters and G.~Pierczynski and P.~Skowron and D.~Stolicki and S.~Szufa and N.~Talmon},  title     = {Participatory budgeting: Data, tools and analysis},
  booktitle = {Proceedings of the International Joint Conference on Artificial Intelligence},
  year      = {2023},
  pages     = {2667--2674}
}

@article{everett1999centrality,
  title={The centrality of groups and classes},
  author={M.~G.~Everett and S.~P.~Borgatti},
  journal={The Journal of Mathematical Sociology},
  volume={23},
  number={3},
  pages={181--201},
  year={1999},
  publisher={Taylor \& Francis}
}

@inproceedings{angriman2020group,
  title={Group centrality maximization for large-scale graphs},
  author={E.~Angriman and A.~van der Grinten and A.~Bojchevski and D.~Z{\"u}gner and S.~G{\"u}nnemann and H.~Meyerhenke},
  booktitle={Proceedings of the Workshop on Algorithm Engineering and Experiments},
  pages={56--69},
  year={2020}
}

@article{holzman2013impartial,
  title={Impartial nominations for a prize},
  author={R.~Holzman and H.~Moulin},
  journal={Econometrica},
  volume={81},
  number={1},
  pages={173--196},
  year={2013},
  publisher={Wiley Online Library}
}

@book{behrens2014principles,
  title={The principles of LiquidFeedback},
  author={J.~Behrens and A.~Kistner and A.~Nitsche and B.~Swierczek},
  year={2014},
  publisher={Interacktive Demokratie}
}

@Article{PetPieSko-2021-EqualSharesPB,
  author   = {D.~Peters and G.~Pierczy{\'n}ski and P.~Skowron},
  journal  = {Advances in Neural Information Processing Systems},
  title    = {Proportional participatory budgeting with additive utilities},
  year     = {2021},
  pages    = {12726--12737},
  volume   = {34}
}

@Article{GirNew-2002-CommunityDetection,
  author    = {M.~Girvan and M.~E.~J.~Newman},
  journal   = {Proceedings of the National Academy of Sciences},
  title     = {Community structure in social and biological networks},
  year      = {2002},
  number    = {12},
  pages     = {7821--7826},
  volume    = {99},
  publisher = {National Acad Sciences}
}

@InProceedings{AdaGla-2005-PolBlogs,
  author    = {L.~A.~Adamic and N.~Glance},
  booktitle = {Proceedings of the Workshop on the Weblogging Ecosystem},
  title     = {The political blogosphere and the 2004 {US} election: divided they blog},
  year      = {2005},
  pages     = {36--43}
}

@article{jin2021survey,
  title={A survey of community detection approaches: From statistical modeling to deep learning},
  author={D.~Jin and Z.~Yu and P.~Jiao and S.~Pan and D.~He and J.~Wu and S.~Y.~Philip and W.~Zhang},
  journal={IEEE Transactions on Knowledge and Data Engineering},
  volume={35},
  number={2},
  pages={1149--1170},
  year={2021},
  publisher={IEEE}
}

@book{langville2006google,
  title={Google's PageRank and beyond: The science of search engine rankings},
  author={A.~N.~Langville and C.~D.~Meyer},
  year={2006},
  publisher={Princeton university press}
}

@inproceedings{mavroforakis2015absorbing,
  title={Absorbing random-walk centrality: Theory and algorithms},
  author={C.~Mavroforakis and M.~Mathioudakis and A.~Gionis},
  booktitle={Proceedings of the IEEE International Conference on Data Mining},
  pages={901--906},
  year={2015}
}

@inproceedings{zhou2022strategyproof,
  title={Strategyproof Mechanisms for Group-Fair Facility Location Problems},
  author={H.~Zhou and M.~Li and H.~Chan},
  booktitle={Proceedings of the International Joint Conference on Artificial Intelligence},
  pages={613--619},
  year={2022}
}

@article{chunaev2020community,
  title={Community detection in node-attributed social networks: a survey},
  author={P.~Chunaev},
  journal={Computer Science Review},
  volume={37},
  pages={100286},
  year={2020},
  publisher={Elsevier}
}

@inproceedings{lam2024proportional,
  title={Proportional fairness in obnoxious facility location},
  author={A.~Lam and H.~Aziz and B.~Li and F.~Ramezani and T.~Walsh},
  booktitle={Proceedings of the International Conference on Autonomous Agents and MultiAgent Systems},
  pages={1075--1083},
  year={2024}
}

@article{blanco2023fairness,
  title={Fairness in maximal covering location problems},
  author={V.~Blanco and R.~G{\'a}zquez},
  journal={Computers \& Operations Research},
  volume={157},
  pages={106287},
  year={2023},
  publisher={Elsevier}
}

@article{bothorel2015clustering,
  title={Clustering attributed graphs: models, measures and methods},
  author={C.~Bothorel and J.~D.~Cruz and M.~Magnani and B.~Micenkova},
  journal={Network Science},
  volume={3},
  number={3},
  pages={408--444},
  year={2015},
  publisher={Cambridge University Press}
}

@misc{kose2021fairness,
  author        = {O.~D.~K{\"o}se and Y.~Shen},
  title         = {Fairness-aware node representation learning},
  year          = {2021},
  eprint        = {2106.05391v1},
  archivePrefix = {arXiv},
  primaryClass  = {cs.LG},
	howpublished={arXiv:2106.05391}
}

@article{shi2024label,
  title={Label deconvolution for node representation learning on large-scale attributed graphs against learning bias},
  author={Z.~Shi and J.~Wang and F.~Lu and H.~Chen and D.~Lian and Z.~Wang and J.~Ye and F.~Wu},
  journal={IEEE Transactions on Pattern Analysis and Machine Intelligence},
  year={2024},
  publisher={IEEE}
}

@Article{Gil-1959-RandomGraphs,
  author   = {E.~N.~Gilbert},
  journal  = {The Annals of Mathematical Statistics},
  title    = {Random graphs},
  year     = {1959},
  number   = {4},
  pages    = {1141--1144},
  volume   = {30}
}

@Article{ErdRen-1959-RandomGraphs,
  author   = {P.~Erdős and A.~Rényi},
  journal  = {Publicationes Mathematicae Debrecen},
  title    = {On random graphs},
  year     = {1959},
  number   = {6},
  pages    = {290--297}
}

@inproceedings{revel2025representative,
  title={Representative ranking for deliberation in the public sphere},
  author={M.~Revel and S.~Milli and T.~Lu and J.~Watson-Daniels and M.~Nickel},
  booktitle={Proceedings of the International Conference on Machine Learning},
  year={2025}
}

@article{sanchez2026proportional,
  title={Proportional justified representation},
  author={L.~S{\'a}nchez-Fern{\'a}ndez and E.~Elkind and M.~Lackner and N.~Fern{\'a}ndez and J.~Fisteus and P.~B.~Val and P.~Skowron},
  journal={Artificial Intelligence},
  pages={104503},
  year={2026},
  publisher={Elsevier}
}

@inproceedings{tsioutsiouliklis2021fairness,
  title={Fairness-aware {P}age{R}ank},
  author={S.~Tsioutsiouliklis and E.~Pitoura and P.~Tsaparas and I.~Kleftakis and N.~Mamoulis},
  booktitle={Proceedings of the Web Conference},
  pages={3815--3826},
  year={2021}
}

@inproceedings{tsang2019groupfairness,
  author = {A.~Tsang and B.~Wilder and E.~Rice and M.~Tambe and Y.~Zick},
  booktitle = {Proceedings of the International Joint Conference on Artificial Intelligence},
  title = {Group-fairness in influence maximization},
  year = {2019},
  pages = {5997--6005}
}

@inproceedings{stoica2020seeding,
  title={Seeding network influence in biased networks and the benefits of diversity},
  author={A.-A.~Stoica and J.~X.~Han and A.~Chaintreau},
  booktitle={Proceedings of the Web Conference},
  pages={2089--2098},
  year={2020}
}

@inproceedings{anwar2021balanced,
  title={Balanced influence maximization in the presence of homophily},
  author={M.~S.~Anwar and M.~Saveski and D.~Roy},
  booktitle={Proceedings of the ACM International Conference on Web Search and Data Mining},
  pages={175--183},
  year={2021}
}

@article{saxena2024fairsna,
  title={FairSNA: Algorithmic fairness in social network analysis},
  author={A.~Saxena and G.~Fletcher and M.~Pechenizkiy},
  journal={ACM Computing Surveys},
  volume={56},
  number={8},
  pages={1--45},
  year={2024}
}

@inproceedings{kilbertus2018blind,
  title={Blind justice: Fairness with encrypted sensitive attributes},
  author={N.~Kilbertus and A.~Gasc{\'o}n and M.~Kusner and M.~Veale and K.~Gummadi and A.~Weller},
  booktitle={Proceedings of the International Conference on Machine Learning},
  pages={2630--2639},
  year={2018}
}

@article{van2023using,
  title={Using sensitive data to prevent discrimination by artificial intelligence: Does the GDPR need a new exception?},
  author={M.~Van Bekkum and F.~Z.~Borgesius},
  journal={Computer Law \& Security Review},
  volume={48},
  pages={105770},
  year={2023}
}

@incollection{koschutzki2005centrality,
  title={Centrality indices},
  author={D.~Kosch{\"u}tzki and K.~A.~Lehmann and L.~Peeters and S.~Richter and D.~Tenfelde-Podehl and O.~Zlotowski},
  booktitle={Network Analysis: Methodological Foundations},
  pages={16--61},
  year={2005}
}

@misc{hardt2015google,
  title={Google {V}otes: A liquid democracy experiment on a corporate social network},
  author={S.~Hardt and L.~C.~Lopes},
  year={2015},
  howpublished={Technical Disclosure Commons}
}

@inproceedings{lederer2024squared,
  title={The Squared Kemeny Rule for Averaging Rankings},
  author={P.~Lederer and D.~Peters and T.~\Was},
  booktitle={Proceedings of the ACM Conference on Economics and Computation},
  pages={755},
  year={2024}
}

@inproceedings{papasotiropoulosfairness,
  title={Fairness in the Multi-Secretary Problem},
  author={G.~Papasotiropoulos and Z.~Pishbin},
  booktitle={Proceedings of the AAAI Conference on Artificial Intelligence},
  pages={17188--17196},
  year={2026}
}

@inproceedings{fish2023generative,
  author       = {S.~Fish and P.~G{\"{o}}lz and D.~C.~Parkes and A.~D.~Procaccia and G.~Rusak and I.~Shapira and M.~W{\"{u}}thrich},
  title        = {Generative social choice},
  booktitle    = {Proceedings of the {ACM} Conference on Economics and Computation},
  pages        = {985},
  year         = {2024}}

@inproceedings{tang2014tim_plus,
  title={Influence maximization: Near-optimal time complexity meets practical efficiency},
  author={Y.~Tang and X.~Xiao and Y.~Shi},
  booktitle={Proceedings of the ACM International Conference on Management of Data},
  pages={75--86},
  year={2014}
}

@inproceedings{goyal2011celf,
  title={Celf++ optimizing the greedy algorithm for influence maximization in social networks},
  author={A.~Goyal and W.~Lu and L.~Lakshmanan},
  booktitle={Proceedings of the International Conference Companion on World Wide Web},
  pages={47--48},
  year={2011}
}

@inproceedings{arora2017debunking,
  title={Debunking the myths of influence maximization: An in-depth benchmarking study},
  author={A.~Arora and S.~Galhotra and S.~Ranu},
  booktitle={Proceedings of the ACM International Conference on Management of Data},
  pages={651--666},
  year={2017}
}

@inproceedings{ashurst2023fairness,
  title={Fairness without demographic data: A survey of approaches},
  author={C.~Ashurst and A.~Weller},
  booktitle={Proceedings of the ACM Conference on Equity and Access in Algorithms, Mechanisms, and Optimization},
  pages={1--12},
  year={2023}
}

	\clearpage
\appendix
\section{Proofs Omitted from the Main Text}

\label{app:proofs}

\subsection*{Violation of Clique-Entitlement by \toprank}
Below we present a graph that demonstrates that \toprank violates Clique-Entitlement, for $k=3$. The graph consists of 2 components, one of size $2$ that is strongly connected, say $G_1,$ and one of size $4$ that is only weakly connected, say $G_2$.

\begin{figure}[h!]
\centering
\begin{tikzpicture}[x=3cm,y=3cm] 
	\node[e4c node] (1) at (-0.40, 0.25) {}; 
	\node[e4c node] (2) at (0.10, 0.25) {}; 
	\node[e4c node,selected] (3) at (0.50, 0.25) {}; 
	\node[e4c node,selected] (4) at (0.75, 0.50) {}; 
	\node[e4c node,selected] (5) at (1.00, 0.25) {}; 
	\node[e4c node] (6) at (0.75, 0) {}; 
	
	\path[e4c path]
	(1) edge[e4c edge, bend left=15]  (2)
	(2) edge[e4c edge, bend left=15]  (1)
	(3) edge[e4c edge, bend left=15]  (4)
	(4) edge[e4c edge, bend left=15]  (3)
	(3) edge[e4c edge, bend left=15]  (5)
	(5) edge[e4c edge, bend left=15]  (3)
	(4) edge[e4c edge, bend left=15]  (5)
	(5) edge[e4c edge, bend left=15]  (4)
	(6) edge[e4c edge]  (3)
	(6) edge[e4c edge]  (4)
	(6) edge[e4c edge]  (5)
	;
	
\end{tikzpicture}
\label{fig:toprank_violates}
\end{figure}
\noindent Under the examined axiom, the nodes from $G_1$ are entitled to $1$ representative in the selected committee. However, the three nodes of maximal PageRank and Katz centrality (nodes with red double lines) all belong to $G_2$. As a result, no node from $G_1$ will be selected under \toprank. The same applies to \topkatz. \hfill \qedhere

\subsection*{Satisfaction of Group-Entitlement by \mesrank}

For every committee election $(V, C, \mu)$ induced by a graph $G$, as described in \Cref{sec:rules}, we define the underlying approval-based election $(V', C', \mu')$, where $\mu_i' : C \to \{0,1\}$. The sets of candidates and voters in the approval-based election are identical copies of those in the original one, i.e., $V'=C'=V=C,$ and, for every pair of nodes $(v,u)$, voter $v$ approves candidate $u$ if and only if there is a path from $v$ to $u$ in $G$.

We show that if an outcome satisfies PJR in the underlying approval-based election it also satisfies Group-Entitlement in the original election.
PJR (for approval-based elections) is defined as follows: A committee $W$ satisfies PJR if for every group $S \subseteq V$ such that $|S| \ge \ell \cdot \tfrac{n}{k}$ and $|\bigcap_{u \in S} A_u | \ge \ell,$
the committee $W$ contains at least $\ell$ candidates from  
$\bigcup_{u \in S} A_u$, where $A_u$ corresponds to the approval set of voter $u$~\citep{sanchez2026proportional}.
For contradiction, assume that a committee $W$ satisfies PJR in the underlying approval-based election but not Group-Entitlement in the original. Then, there exists a subset of nodes $S\subseteq V$ with
$|S|\geq \ell \cdot \nicefrac{n}{k}$
and $|\bigcap_{u\in S}\Succ(u)| \geq \ell$ such that
$|\Succ(S) \cap W| < \ell.$
The fact that $|\bigcap_{u\in S}\Succ(u)| \geq \ell$ means that in the underlying approval-based election $|\bigcap_{u\in S}A_u| \geq \ell$.
The fact that PJR is satisfied means that $W$ contains at least $\ell$ candidates from  
$\bigcup_{u \in S} A_u$. But for every candidate in $\bigcup_{i \in S} A_i$ in the underlying approval-based election, its corresponding candidate in the original election belongs to $\Succ(S)$, hence $|\Succ(S) \cap W| \geq \ell$.
Now it remains to show that the outcome of \mesrank (and \meskatz) satisfies PJR. Note that its outcome is priceable due to the fact that it was obtained by running MES in a (cardinal) committee election, and such outcomes are priceable \cite[Section 3.3]{PetPieSko-2021-EqualSharesPB}.
From~\citet{Peters:Skowron:2020} (Proposition 1 therein) we know that every priceable outcome satisfies PJR in the underlying approval-based election.
\hfill 
\qedhere

\section{Additional Experimental Results}\label{sec:appendix-experiments}
\label{app:experimental_results}

In \Cref{fig:polblogs:plots:appendix} we present plots analogous to those in \Cref{fig:polblogs:plots_PR}, but for $k=20$, instead of $k=10$.

\begin{figure*}[t]
\centering
\input{figures/polblog_plots_appendix.tex}
\caption{		
Results of our experiments on Political Blogs Network for $k=20$.
Each column presents the values of a different statistic
depending on the probability of removing blogs from one side ($x$-axis).
In the first column, we show the average
fraction of minority-labeled nodes in the outputs of the rules
(in this column we also present the fraction of minority-labeled nodes
among all nodes in the network---grey dashed line).
In the second and third columns, we plot the
average summed values of PageRank and Katz centrality, respectively,
among the selected nodes.
In the fourth column,
we present the average number of infected nodes
in the independent cascades model,
when the seed is an output of a rule.
Top row shows the values for \toprank, \mesrank, and \bosrank,
while the bottom one for \topkatz, \meskatz, and \boskatz.
The values for the remaining methods are identical in both rows.
Vertical bars denote 95\% confidence intervals.}
\label{fig:polblogs:plots:appendix}
\end{figure*}

\end{document}